\documentclass[apjl]{emulateapj}
\usepackage{psfig,amsfonts,amsmath,graphicx,natbib,apjfonts}
\def\nod{\nodata}
\def\lsr{LSR\,1835+32}
\def\vb{VB\,10}
\def\tvlm{TVLM\,513-46546}
\def\swift{{\it Swift}}

\def\prince{1}
\def\ociw{2}
\def\ucb{3}
\def\udel{4}
\def\noao{5}
\def\mcgill{6}
\def\steward{7}
\def\iac{8}
\def\ucf{9}
\def\rice{10}
\def\iaa{11}

\begin{document}

\title{Simultaneous Multi-Wavelength Observations of Magnetic Activity
in Ultracool Dwarfs. II.  Mixed Trends in VB\,10 and LSR\,1835+32 and
the Possible Role of Rotation}

\author{
E.~Berger\altaffilmark{\prince,}\altaffilmark{\ociw},
G.~Basri\altaffilmark{\ucb},
J.~E.~Gizis\altaffilmark{\udel},
M.~S.~Giampapa\altaffilmark{\noao},
R.~E.~Rutledge\altaffilmark{\mcgill},
J.~Liebert\altaffilmark{\steward},
E.~Mart{\'{\i}}n\altaffilmark{\iac,}\altaffilmark{\ucf},
T.~A.~Fleming\altaffilmark{\steward},
C.~M.~Johns-Krull\altaffilmark{\rice},
N.~Phan-Bao\altaffilmark{\iaa},
W.~H.~Sherry\altaffilmark{\noao}
}

\altaffiltext{\prince}{Princeton University Observatory,
Peyton Hall, Ivy Lane, Princeton, NJ 08544}

\altaffiltext{\ociw}{Observatories of the Carnegie Institution
of Washington, 813 Santa Barbara Street, Pasadena, CA 91101}

\altaffiltext{\ucb}{Astronomy Department, University of
California, Berkeley, CA 94720}

\altaffiltext{\udel}{Department of Physics and Astronomy,
University of Delaware, Newark, DE 19716}

\altaffiltext{\noao}{National Solar Observatory, National Optical
Astronomy Observatories, Tucson, AZ 85726}

\altaffiltext{\mcgill}{Department of Physics, McGill University,
3600 University Street, Montreal, QC H3A 2T8, Canada}

\altaffiltext{\steward}{Department of Astronomy and Steward
Observatory, University of Arizona, 933 North Cherry Avenue,
Tucson, AZ 85721}

\altaffiltext{\iac}{Instituto de Astrof{\'{\i}}sica de Canarias,
C/ V{\'{\i}}a L\'actea s/n, E-38200 La Laguna, Tenerife, Spain}

\altaffiltext{\ucf}{University of Central Florida, Department of
Physics, PO Box 162385, Orlando, FL 32816}

\altaffiltext{\rice}{Department of Physics and Astronomy, Rice
University, 6100 Main Street, MS-61 Houston, TX 77005}

\altaffiltext{\iaa}{Institute of Astronomy and Astrophysics, Academia
Sinica, PO Box 23-141, Taipei 106, Taiwan, ROC}

\begin{abstract} As part of our on-going investigation of magnetic
activity in ultracool dwarfs we present simultaneous radio, X-ray, UV,
and optical observations of \lsr\ (M8.5), and simultaneous X-ray and
UV observations of \vb\ (M8), both with a duration of about 9 hours.
\lsr\ exhibits persistent radio emission and H$\alpha$ variability on
timescales of $\sim 0.5-2$ hr.  The detected UV flux is consistent
with photospheric emission, and no X-ray emission is detected to a
deep limit of $L_X/L_{\rm bol}\lesssim 10^{-5.7}$.  The H$\alpha$ and
radio emission are temporally uncorrelated, and the ratio of radio to
X-ray luminosity exceeds the correlation seen in F--M6 stars by
$\gtrsim\!2\times 10^4$.  Similarly, $L_{\rm H\alpha}/L_X\gtrsim 10$
is at least 30 times larger than in early M dwarfs, and eliminates
coronal emission as the source of chromospheric heating.  The lack of
radio variability during four rotations of \lsr\ requires a uniform
stellar-scale field of $\sim 10$ G, and indicates that the H$\alpha$
variability is dominated by much smaller scales, $\lesssim 10\%$ of
the chromospheric volume.  \vb, on the other hand, shows correlated
flaring and quiescent X-ray and UV emission, similar to the behavior
of early M dwarfs.  Delayed and densely-sampled optical spectra
exhibit a similar range of variability amplitudes and timescales to
those seen in the X-rays and UV, with $L_{\rm H\alpha}/L_X\sim 1$.
Along with our previous observations of the M8.5 dwarf \tvlm\ we
conclude that late M dwarfs exhibit a mix of activity patterns, which
points to a transition in the structure and heating of the outer
atmosphere by large-scale magnetic fields.  We find that rotation may
play a role in generating the fields as evidenced by a tentative
correlation between radio activity and rotation velocity.  The X-ray
emission, however, shows evidence for super-saturation at $v{\rm
sin}i\gtrsim 25$ km s$^{-1}$, which could be the result of secondary
effects such as inefficient heating or centrifugal stripping of
extended coronal loops.  These effects may underlie the severe
violation of the radio/X-ray correlation in ultracool dwarfs.
Upcoming observations of L dwarfs will reveal whether these trends
continue in substellar objects.  \end{abstract}
 

\section{Introduction}
\label{sec:intro}

Over the past several decades, observations of magnetic activity in
stars of spectral type F--M have uncovered a variety of correlations
between the different activity indicators, as well as between the
level of activity and properties such as stellar rotation and age.
Coronal X-ray emission, transition region UV emission, chromospheric
H$\alpha$ emission, and non-thermal radio emission increase with both
rotation and youth (e.g., \citealt{kra67,pgr+81,sis+88}), and are
temporally and energetically correlated in quiescence and during
flares (e.g., \citealt{cra82,gb93,bg94,hgr96}).  These observations
have led to a general paradigm of magnetic field amplification at the
shearing interface between the radiative and convective zones -- the
so-called $\alpha\Omega$ dynamo \citep{par55}.  This dynamo operates
through a combination of stretching by differential rotation
($\Omega$) and twisting by convective motions ($\alpha$).

In quiescence, the dynamo-generated fields provide a heating source
for the chromospheres and coronae primarily through magnetic waves and
field dissipation on small scales (e.g., \citealt{nu96,apr01}).  The
interplay between the chromosphere, transition region, and corona is
not fully understood, but is known to involve a complex combination of
radiation, conduction, and mass flows.  In dMe stars it has been
proposed that the transition region and chromosphere are instead
heated by coronal X-rays, leading to the observed typical luminosity
ratios of $L_{\rm CIV}/L_X\sim 10^{-1.5}$ and $L_{\rm H\alpha}/L_X\sim
10^{-0.5}$, respectively (e.g., \citealt{cra82,hj03}).  In addition to
quiescent emission, sudden and large-scale dissipation of the field
through large-scale magnetic reconnection may lead to particle
acceleration and evaporation of the lower atmosphere into the
chromosphere, transition region, and corona, giving rise to correlated
radio, X-ray, UV, and H$\alpha$ flares (e.g., \citealt{neu68}).

The level of activity increases with faster rotation, but eventually
saturates at $L_X/L_{\rm bol}\sim 10^{-3}$ and $L_{\rm H\alpha}/L_{\rm
bol}\approx 10^{-3.5}$ for rotation periods of $\lesssim 3$ d in F--M
stars.  It remains unclear whether this trend reflects a saturation of
the dynamo itself or secondary centrifugal effects such as coronal
stripping or sweeping of the field toward the poles (e.g.,
\citealt{vil84,ju99,ssv01}).

These activity trends continue to hold even beyond spectral type $\sim
{\rm M3}$, where the stellar interiors become fully convective and the
$\alpha\Omega$ dynamo can no longer operate.  Indeed, the level of
X-ray and H$\alpha$ activity peaks in mid M dwarfs at saturated levels
\citep{vw87,gmr+00,mb03,whw+04}.  This suggests that whatever dynamo
mechanism operates in these low mass stars, it is already present in
higher mass objects, and becomes increasingly dominant beyond spectral
type M3.  However, the level of X-ray and H$\alpha$ activity drops
precipitously beyond spectral type M7, reaching mostly non-detectable
levels by spectral type L5.  This decrease is accompanied by a clear
transition from quiescent emission to flares in the few percent of
active objects \citep{rkg+99,gmr+00,rbm+00,lkc+03,whw+04}.  These
trends point to a change in the dynamo mechanism, the field
configuration, and/or the field dissipation process.

Radio observations, however, have uncovered a substantial fraction of
active late M and L dwarfs ($\sim 10\%$ or higher; \citealt{ber06}),
which exhibit both quiescent and flaring emission
\citep{bbb+01,ber02,brr+05,bp05,ber06,ohb+06,adh+07,pol+07,hbl+07,aob+07,bgg+07}.
Unlike the trend in H$\alpha$ and X-rays, the level of radio activity
appears to {\it increase} with later spectral type
\citep{ber02,ber06}.  The radio emission is orders of magnitude
brighter than expected based on the radio/X-ray correlations that are
observed in stars down to spectral type M6, and requires magnetic
fields of $\sim 0.1-3$ kG with order unity filling factor sustained
over timescales of at least several years \citep{brr+05,ber06,bgg+07}.
Thus, contrary to evidence from X-rays and H$\alpha$, a substantial
fraction of ultracool dwarfs continue to generate and dissipate
magnetic fields.

In order to investigate the field generation and dissipation in
detail, we have recently initiated a program of simultaneous,
multi-wavelength observations of ultracool dwarfs.  Such observations
are required to trace the temporal evolution of flares across the
corona, transition region, and chromosphere, and to study the relation
between particle acceleration and heating, particularly in the context
of the known correlations in F--M stars.  In a previous paper
(\citealt{bgg+07}; hereafter Paper I) we presented observations of the
M8.5 rapid rotator \tvlm, which exhibited a wide range of temporally
uncorrelated emission in the radio, H$\alpha$, and X-rays.  These
included quiescent radio emission from a large-scale field, radio
flares from a tangled field component with $B\approx 3$ kG, and
periodic H$\alpha$ emission matching the stellar rotation period with
an inferred hot spot covering about $50\%$ of the stellar photosphere.
In addition, the quiescent H$\alpha$ emission exceeded the X-ray
luminosity by about a factor of two, likely ruling out coronal X-ray
emission as the source of chromospheric heating.

Here we present observations of the M8 and M8.5 dwarfs \vb\ and \lsr,
both of which are known to exhibit magnetic activity.  We find
substantially different behavior in each of these two objects, with
correlated X-ray/UV flares and quiescent emission in \vb, and
uncorrelated radio/H$\alpha$ emission in \lsr.  Along with \tvlm, the
mixed behavior in late M dwarfs thus indicates a transition in the
properties of the magnetic field and its dissipation in this spectral
type range.  We show that rotation may play at least a partial role in
explaining these trends.

\section{Targets and Observations}
\label{sec:obs}

We chose to target the nearby dwarf stars \vb\ and \lsr\ due to
previous detections of activity from these objects in the
optical/UV/X-ray and radio bands, respectively.  \lsr\ (8.5) is
located at a distance of 5.7 pc and has a bolometric luminosity of
$L_{\rm bol}\approx 10^{-3.4}$ L$_\odot$ \citep{rcl+03}.  A previous
$1.7$-hr radio observation revealed persistent emission with
$F_\nu(8.46)=525\pm 15$ $\mu$Jy and a limit of $\lesssim 9\%$ on the
fraction of circular polarization \citep{ber06}.  The rotation
velocity of \lsr\ was not previously known, but we measure it here to
be $v{\rm sin}i=50\pm 5$ km s$^{-1}$ (\S\ref{sec:optical}), similar to
fast rotators such as \tvlm\ and Kelu-1.

\vb\ (M8) is located at a distance of 6.1 pc \citep{tin96}, has
$L_{\rm bol}\approx 10^{-3.34}$ L$_\odot$, and a slow rotation
velocity, $v{\rm sin}i\approx 6.5$ km s$^{-1}$ \citep{mb03}.  An X-ray
flare from \vb\ was previously detected by ROSAT, with a duration of
$\sim 20$ min and an average luminosity, $L_X=(8.4\pm 2.7)\times
10^{26}$ erg s$^{-1}$, or $L_X/L_{\rm bol}\approx 10^{-3.3}$
\citep{fgs00}.  Subsequent Chandra observations revealed quiescent
emission with $L_X/L_{\rm bol}\approx 10^{-4.9}$ \citep{fgg03}.  In
addition, {\it Hubble Space Telescope} observations revealed flaring
and quiescent transition region emission \citep{lwb+95,hj03}.  \vb\ is
also known to produce H$\alpha$ emission with equivalent widths of
$\sim 2-6$ \AA\ reported in the literature \citep{mdb+99,mb03,rb07},
or $L_{\rm H\alpha}/L_{\rm bol}\approx 10^{-4.4}$.  No radio emission
has been detected to date, with a limit of $\lesssim 80$ $\mu$Jy at
8.5 GHz \citep{kll99}.  Finally, \citet{rb07} recently estimated the
surface magnetic field to be $Bf\sim 1.3$ kG, where $f$ is the field
filling factor.

The simultaneous observations of \lsr\ presented here were conducted
on 2007 May 3 for a total of 8.7 hr in the radio (06:40--15:20 UT),
8.4 hr in the X-rays (07:16--15:43 UT), and 5.4 hr in the optical
(09:42--15:04 UT).  Observations with the \swift/UVOT took place
intermittently between 08:09 and 16:31 UT with a total on-source
exposure time of 9.4 ks.  

Observations of \vb\ were conducted on 2007 July 1 for a total of 8.9
hr in the X-rays (10:29--19:21 UT), with intermittent UV coverage
between 11:08 and 19:15 UT for a total of 8.4 ks.  Optical
spectroscopy was obtained only two weeks later, on 2007 July 15
(08:10--13:33 UT).

\subsection{Radio} 
\label{sec:rad}

Very Large Array\footnotemark\footnotetext{The VLA is operated by the
National Radio Astronomy Observatory, a facility of the National
Science Foundation operated under cooperative agreement by Associated
Universities, Inc.} observations of \lsr\ were obtained at a frequency
of 8.46 GHz in the standard continuum mode with $2\times 50$ MHz
contiguous bands.  Scans of 295 s on source were interleaved with 50 s
scans on the phase calibrator J1850+284. The flux density scale was
determined using the extragalactic source 3C\,286 (J1331+305).  Data
reduction and analysis follow the procedure outlined in Paper I.  The
total intensity and circular polarization light curves are shown in
Figure~\ref{fig:lsrall}.

\subsection{X-Rays}
\label{sec:xrays}

Observations were performed with the Chandra/ACIS-S3
backside-illuminated chip, with both \lsr\ and \vb\ offset from the
on-axis focal point by $3.17'$.  A total of 28.33 and 29.29 ks were
obtained, respectively.  The data were analyzed using CIAO version
3.4, and counts were extracted in a $2''$ radius circle centered on
the position of each source.

For \lsr\ we established a bore-sight correction based on one source
in common between the VLA and Chandra images.  The derived correction
is $\delta{\rm RA}=-1.7\pm 0.7''$ and $\delta{\rm DEC}=0.2\pm 0.5''$.
Within the $2''$ aperture centered on the position of \lsr\ we find
only 2 counts, with 1.5 counts expected from the background as
determined from annuli centered on the source position.  Thus, the
resulting upper limit is about 7 counts ($95\%$ confidence level).
Using an energy conversion factor of $1\,{\rm cps}=3.4\times 10^{-12}$
erg cm$^{-2}$ s$^{-1}$ (appropriate for a 1 keV Raymond-Smith plasma
model in the $0.2-2$ keV range) we find $F_X<8.4\times 10^{-16}$ erg
cm$^{-2}$ s$^{-1}$, or $L_X/L_{\rm bol}<10^{-5.7}$, one of the
faintest limits to date for any ultracool dwarf.

In the observation of \vb\ we detect a total of 60 counts within the
$2''$ extraction aperture; 2 counts are expected from the background.
Of the 60 detected counts, 59 are in the energy range $0.2-2$ keV,
similar to what has been found for other M dwarfs (including \tvlm:
Paper I).  Using this energy range we construct light curves and
spectra.  The light curve is shown at various time binnings in
Figure~\ref{fig:vb10all}, and is composed of a bright flare with a
duration of $\sim 3$ hr, followed by about 4 hr of low-level quiescent
emission, and finally a fainter flare during the final hour of the
observation.  Using the time range 14:30--18:00 UT to represent the
quiescent component we find a total of 10 counts, or a count rate of
$\approx 6.2\times 10^{-4}$ s$^{-1}$.

\begin{figure}
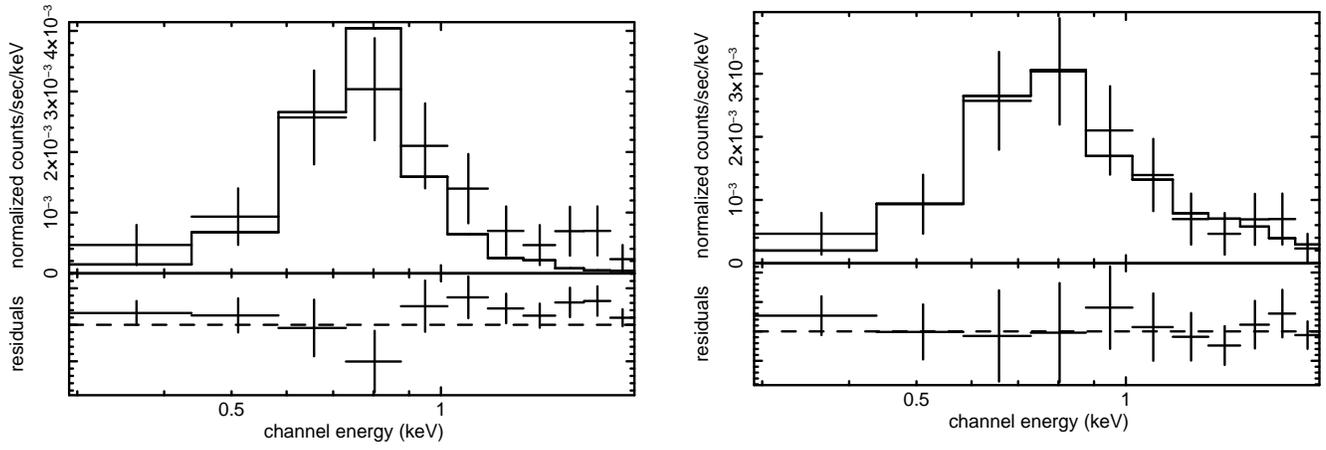

\begin{minipage}{9cm}
\includegraphics[angle=270,width=3.65in]{fig3a.ps}
\end{minipage} \hfill
\begin{minipage}{9cm}
\includegraphics[angle=270,width=3.65in]{fig3b.ps}
\end{minipage} 
\caption{X-ray spectrum of \vb\ (data with error bard) fit with {\it
Left:} a single temperature Raymond-Smith model with $kT\approx 0.3$
keV ($\chi^2_r=1.4$ for 9 d.o.f), and {\it Right:} a two-component
Raymond-Smith model with $kT_1\approx 0.26$ keV and $kT_2\approx 1.3$
keV ($\chi^2_r=0.3$ for 7 d.o.f).  The single-temperature model leaves
significant residuals, which are largely eliminated by the addition of
hotter component.
\label{fig:vb10xspec}}
\end{figure}

To determine the flux and plasma temperature we fit the $0.2-2$ keV
spectrum using a Raymond-Smith model.  We find that a single-component
model provides a poor fit to the data, with $\chi^2_r=1.4$ for 9
degrees of freedom (null hypothesis probability of 0.2);
Figure~\ref{fig:vb10xspec}.  The fit parameters are $kT\approx 0.3$
keV, and a normalization of $1.4\times 10^{-5}$.

An improved fit is achieved using two components, with $\chi^2_r=0.3$
for 7 degrees of freedom (null hypothesis probability of 0.94);
Figure~\ref{fig:vb10xspec}.  The best-fit temperatures are $kT_1=
0.26^{+0.06}_{-0.03}$ keV and $kT_2\approx 1.3$ keV (formally,
$kT_2\gtrsim 1$ keV).  The normalizations of the two components are
$(1.35\pm 0.35)\times 10^{-5}$ and $(6.5\pm 2.1)\times 10^{-6}$,
respectively.  The resulting average flux over the full observation is
$F_X=(1.2\pm 0.4)\times 10^{-14}$ erg cm$^{-2}$ s$^{-1}$, or
$L_X/L_{\rm bol}\approx 10^{-4.5}$.

Using the same model for the quiescent component, we find $F_X\approx
3.7\times 10^{-15}$ erg cm$^{-2}$ s$^{-1}$, or $L_X/L_{\rm bol}\approx
10^{-5.0}$, in excellent agreement with the value found previously by
\citet{fgg03}.  The peak count rate of the first flare is about
$5\times 10^{-3}$ s$^{-1}$, corresponding to $F_X\approx 3.2\times
10^{-14}$ erg cm$^{-2}$ s$^{-1}$ and $L_X/L_{\rm bol}\approx
10^{-4.1}$.  For the second flare, $F_X\approx 2.0\times 10^{-14}$ erg
cm$^{-2}$ s$^{-1}$ and $L_X/L_{\rm bol}\approx 10^{-4.3}$.

\subsection{Optical Spectroscopy} 
\label{sec:optical}

We used\footnotemark\footnotetext{Observations were obtained as part
of program GN-2007A-Q-60.} the Gemini Multi-Object Spectrograph (GMOS;
\citealt{hja+04}) mounted on the Gemini-North 8-m telescope with the
B600 grating set at a central wavelength of 5250 \AA, and with a $1''$
slit.  The individual 300-s exposures were reduced using the {\tt
gemini} package in IRAF (bias subtraction, flat-fielding, and sky
subtraction).  Wavelength calibration was performed using CuAr arc
lamps and air-to-vacuum corrections were applied.  The spectra cover
$3840-6680$ \AA\ at a resolution of about 5 \AA.  For \lsr\ we
obtained a series of sixty exposures, while for \vb\ the data were
obtained non-simultaneously with a total of 61 exposures.  The fast
readout time of $18$ s provides $94\%$ on-source efficiency.

We detect highly variable H$\alpha$ emission in \lsr\
(Figure~\ref{fig:lsrall}).  Spectra with strong H$\alpha$ emission
also exhibit higher-order Balmer lines (H$\beta$--H$\delta$) and
\ion{Ca}{2} H\&K emission.  Sample spectra in the High and Low
emission line states are shown in Figure~\ref{fig:lsrspec}.

\begin{figure}
\centerline{\psfig{file=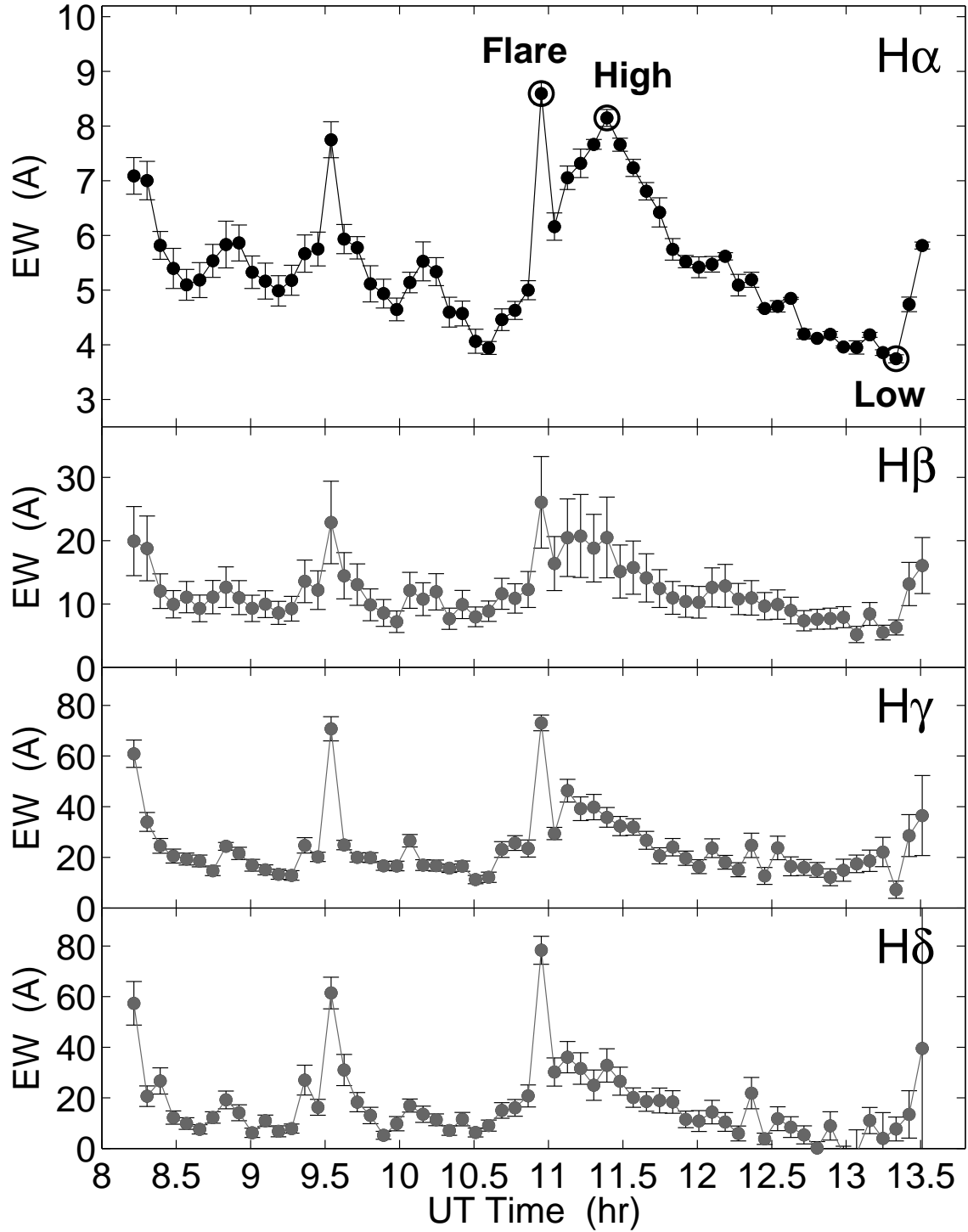,width=6.0in}}
\caption{Light curves of the Balmer line equivalent widths in the
spectra of \vb.  Each spectrum has a duration of 300 s.  The same
behavior is clear in all four lines, with significant variability
ranging in duration from a single exposure to a broad brightening with
a duration of about 2.5 hr.  We designate three spectral states ---
Flare, High, and Low --- whose spectra are shown in
Figure~\ref{fig:vb10spec}.  The flare state is more distinct in the
higher order Balmer lines, suggesting that the plasma densities and
temperatures are higher compared to the High and Low states.
\label{fig:vb10balmer}}
\end{figure}

\begin{figure}
\centerline{\psfig{file=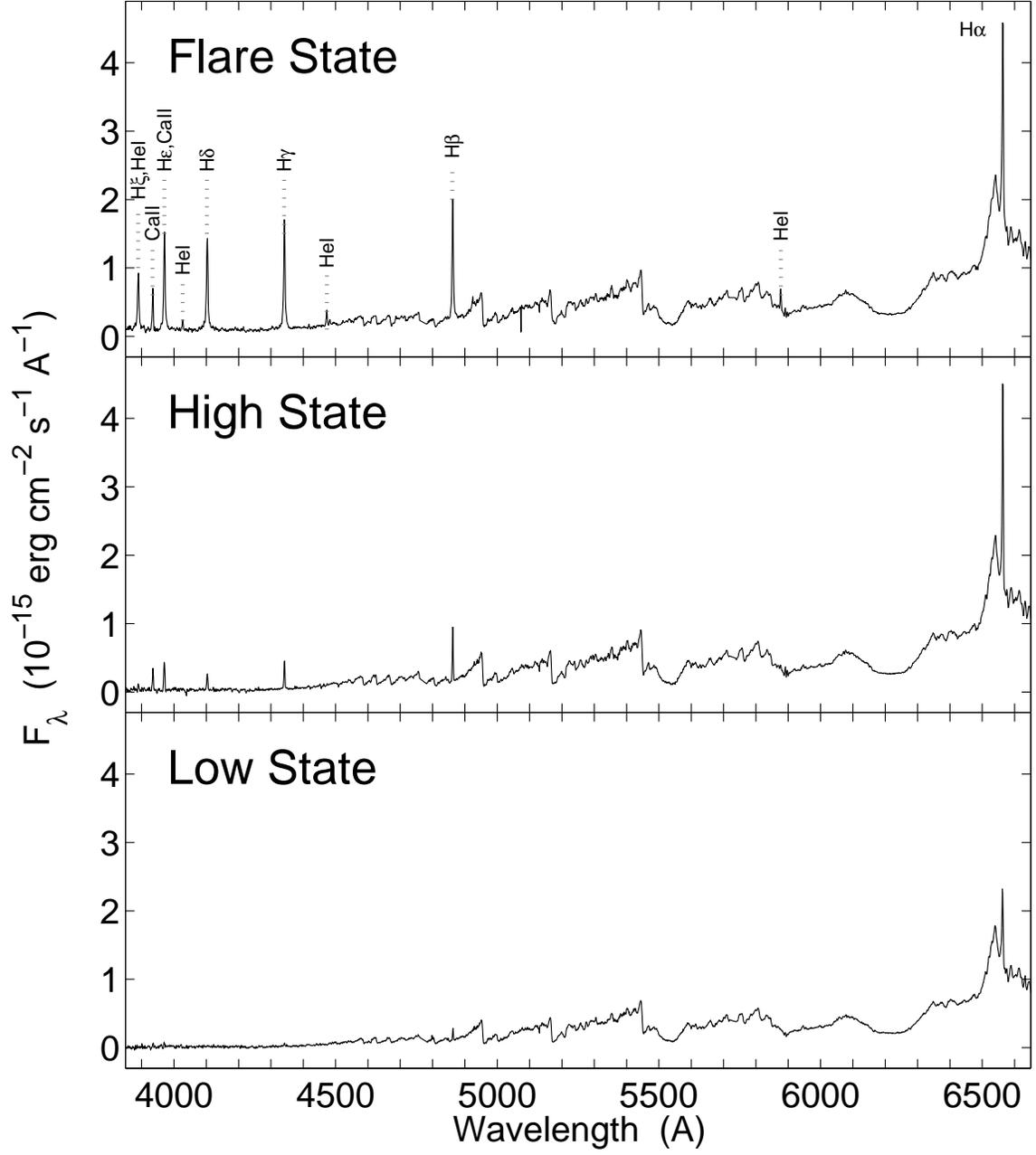,width=6.0in}}
\caption{Spectra of \vb\ in the three states designated in
Figure~\ref{fig:vb10balmer}: Flare, High, and Low.  The flare spectrum
exhibits significantly stronger emission in the higher-order Balmer
lines than the high state spectrum, stronger \ion{Ca}{2} lines, and
\ion{He}{1} lines which are completely absent in the High state
spectrum.  These properties point to higher chromospheric densities 
and temperatures in the impulsive Flare state ($\lesssim 300$ s) than 
during the broad brightenings (High state).  The Low state spectrum 
is simply a weaker version of the High state spectrum.
\label{fig:vb10spec}} 
\end{figure}

The spectrum of \vb\ exhibits strong variability in the Balmer
(H$\alpha$--H$\xi$) and \ion{Ca}{2} H\&K lines;
Figure~\ref{fig:vb10balmer}.  As in the case of \lsr\ we identify High
and Low emission line states (Figure~\ref{fig:vb10spec}), but we also
identify an impulsive Flare state marked by a short duration
($\lesssim 300$ s; Figure~\ref{fig:vb10balmer}) and emission lines of
\ion{He}{1} (Figure~\ref{fig:vb10spec}).

\begin{figure}
\centerline{\psfig{file=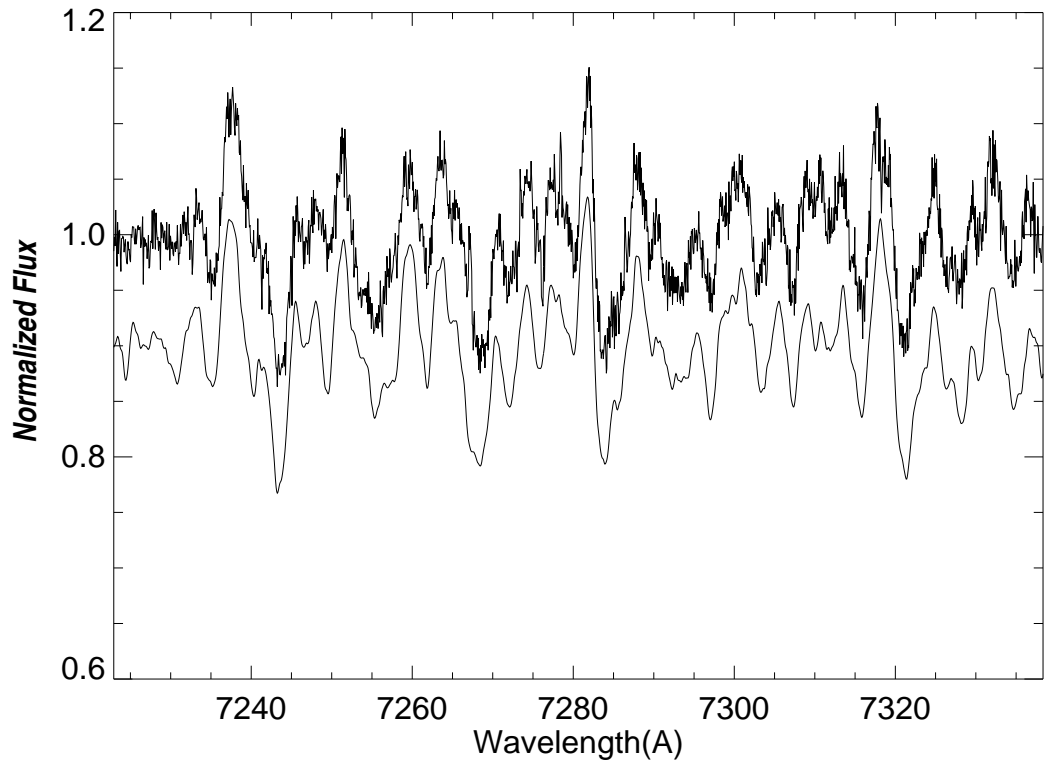,width=6.0in}}
\caption{A portion of our Keck/HIRES echelle spectrum of \lsr\ (thick
line).  A comparison to the rotationally-broadened spectrum of the
slowly-rotating M6 dwarf CN Leo (thin line; shifted downward for 
clarity) indicates a rotation velocity for \lsr\ of $v{\rm sin}i=50
\pm 5$ km s$^{-1}$.
\label{fig:lsrrot}} 
\end{figure}

In addition to the low-resolution Gemini spectra, we obtained high
resolution spectra of \lsr\ to measure its rotation velocity, using
the High Resolution Echelle Spectrometer (HIRES) on the Keck I 10-m
telescope.  The spectroscopic setup and data reduction are detailed in
\citet{mb03} and \citet{rb07}.  Using the slow rotator CN Leo (M6;
$v{\rm sin}i\approx 3$ km s$^{-1}$) as a template, we measure a
rotation velocity for \lsr\ of $v{\rm sin}i=50\pm 5$ km s$^{-1}$, or a
period of about 2.4 hr; Figure~\ref{fig:lsrrot}.

\subsection{Ultraviolet}

\begin{figure}
\centerline{\psfig{file=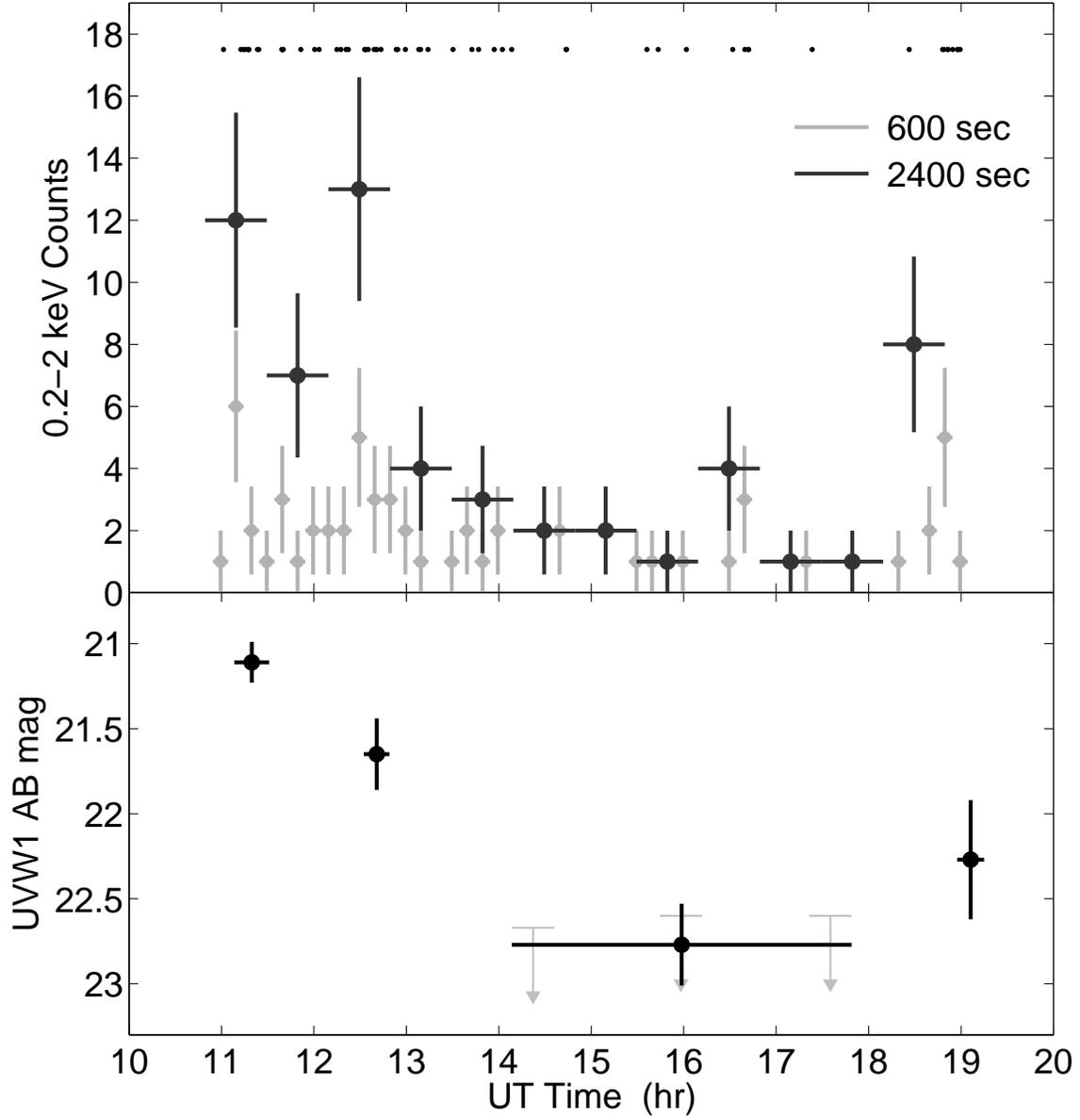,width=6.0in}}
\caption{X-ray and UV observations of \vb.  We show the X-ray photon
arrival times (black dots), as well as light curves with time binning
of 600 and 2400 s.  The light curve is clearly composed of a flare
with a duration of about 2 hrs, followed by quiescent emission and a
second weaker and shorter flare.  The UV light curve clearly tracks
the X-ray emission.
\label{fig:vb10all}}
\end{figure}

Data were obtained with the \swift\ UV/optical telescope in the UVW1
filter ($\lambda_{\rm eff}\approx 2600$ \AA), as a series of 6 images
with exposure times ranging from 560 to 1630 s for \lsr\
(Figure~\ref{fig:lsrall}) and 980 to 1630 s for \vb\
(Table~\ref{tab:vb10uvot} and Figure~\ref{fig:vb10all}).

\begin{figure}
\centerline{\psfig{file=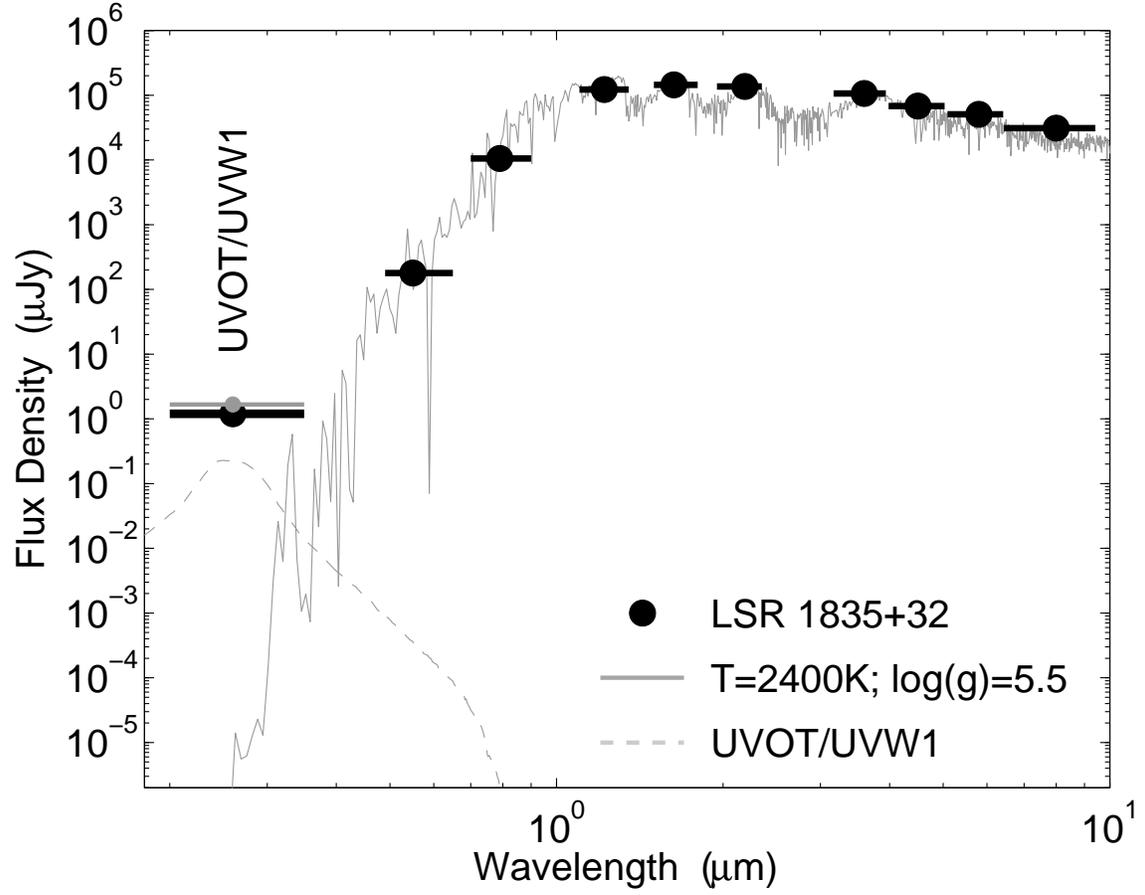,width=6.0in}}
\caption{UV to IR spectral energy distribution of \lsr\ from DSS, 
2MASS, and Spitzer observations.  The gray curve is an AMES-Cond 
atmospheric model of a 2400 K, ${\rm log}g=5.5$ dwarf star, which 
provides an excellent fit to the optical, NIR, and IR data.  
Convolution of the atmospheric model with the UVOT/UVW1 filter 
curve (dashed line) indicates an expected flux (grey square) that 
is in good agreement with the observed flux.
\label{fig:lsrsed}}
\end{figure}

No source is detected at the position of \lsr\ in any of the
individual exposures, but we detect a source at $3.5\sigma$ confidence
level in the combined 9.4 ks image.  Photometry of the combined image
(including a 0.2 mag correction to the standard $5''$ aperture)
reveals a flux of $F_\lambda=(4.7\pm 1.3)\times 10^{-18}$ erg
cm$^{-2}$ s$^{-1}$ \AA$^{-1}$, or $m_{\rm AB}=23.7\pm 0.3$ mag.  To
estimate the expected UV photospheric emission we convolve the
UVOT/UVW1 transmission curve with an AMES-cond
model\footnotemark\footnotetext{The AMES-cond model assumes that all
dust has gravitationally settled out of the atmosphere \citep{aha+01}.
We find that the opposite case of no gravitational settling
(AMES-dusty model) provides essentially the same result in the UV.}
with ${\rm log}(g)=5.5$ and $T_{\rm eff}=2400$ K \citep{aha+01}, 
which provides an excellent fit to the optical and IR data for \lsr\
(Figure~\ref{fig:lsrsed}).  We find an expected photospheric
brightness of $m_{\rm AB}\approx 23.3$ mag, in excellent agreement
with the observed flux.  Thus, no contribution from a quiescent
or flaring transition region is detected.

\vb, on the other hand, is clearly variable in the UV, and is detected
in exposures 1, 2, and 6, with an additional detection when combining
exposures 3, 4, and 5.  A summary of the derived magnitudes, corrected
to the standard $5''$ aperture, is provided in
Table~\ref{tab:vb10uvot}, and the light curve is shown in
Figure~\ref{fig:vb10all}.  Repeating the same analysis as for \lsr, we
find that the minimum observed UV emission exceeds the expected
photospheric emission by about an order of magnitude.  Thus, in \vb\
the flaring and quiescent UV emission are due to an active transition
region.

\section{Multi-Wavelength Emission Properties} 
\label{sec:prop}

We observed \lsr\ and \vb\ across a wide wavelength range that traces
activity in various layers of the outer atmosphere.  The radio
emission is due to particle acceleration by magnetic processes, the
optical emission lines trace the chromospheric plasma, the UV emission
arises in the transition region, and the X-ray thermal emission is
produced in the corona.

\subsection{\lsr} 
\label{sec:lsr}

\begin{figure}
\centerline{\psfig{file=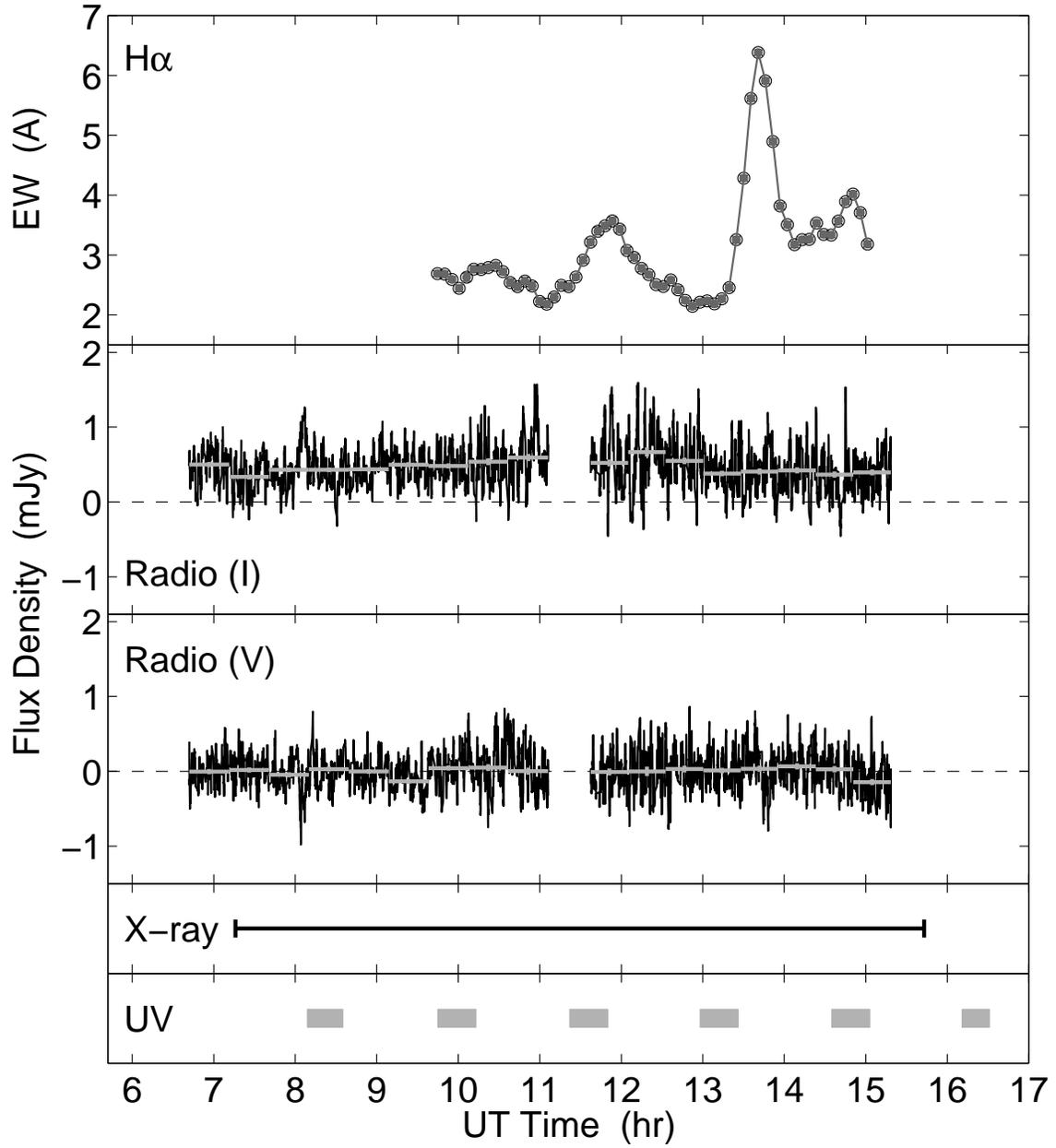,width=6.0in}}
\caption{Radio, H$\alpha$, UV, and X-ray observations of \lsr.  No
X-ray emission is detected, and we find a UV source only in the
combined 9.4 ks observation with $m_{\rm AB}=23.7\pm 0.3$ mag.  The
radio total intensity (I) and circular polarization (V) light curves
(black lines) are shown at 30-sec (black lines) and 30-min (gray
lines) resolution (the intrinsic time resolution is 5 s).  One
possible weak radio flare is detected (08:05 UT), but the overall
level of variability is less than a factor of two.  The H$\alpha$ line
exhibits significant variability on timescales of $0.5-2$ hr.  We find
no correlation between the H$\alpha$ and radio light curves.
\label{fig:lsrall}}
\end{figure}

\begin{figure}
\centerline{\psfig{file=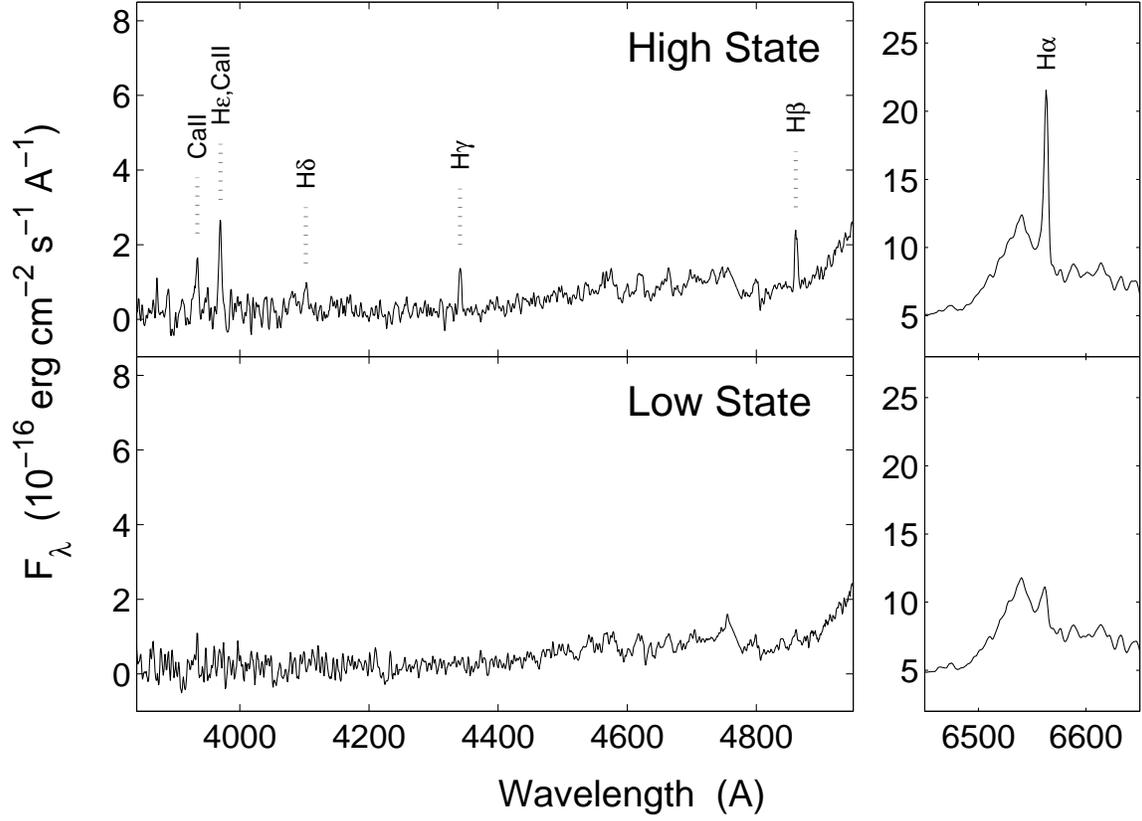,width=6.0in}}
\caption{Spectra of \lsr\ in the High and Low emission lines states,
corresponding, respectively, to the peak (13:40 UT) and lowest point
(13:00 UT) of the H$\alpha$ light curve (Figure~\ref{fig:lsrall}).
The High state is marked by strong Balmer and \ion{Ca}{2} H\&K lines.
\label{fig:lsrspec}} 
\end{figure}

The broad-band emission properties of \lsr\ are summarized in
Figure~\ref{fig:lsrall}.  The radio emission appears to be nearly
constant both in total intensity and circular polarization.  The
average flux density is $F_\nu(8.46)=464\pm 10$ $\mu$Jy, while the
$3\sigma$ upper limit on the fraction of circular polarization is
$r_c<6.5\%$.  For comparison, previous observations of \lsr\ revealed
a flux of $F_\nu(8.46)=525\pm 15$ $\mu$Jy, with an upper limit on the
circular polarization of $r_c<9\%$ \citep{ber06}.  We note the
possible detection of one weak flare (at 08:05 UT) with a peak flux
density of about 1.3 mJy and a duration of about 5 min
(Figure~\ref{fig:lsrall}).  The average flux density during the
putative flare is only $370\pm 105$ $\mu$Jy above the quiescent
emission level, and the average fraction of circular polarization is
$r_c=-50\pm 15\%$.

We therefore conclude that the radio emission from \lsr\ is generally
persistent in origin, with at most a mild variability at the level of
$\sim 20\%$ on hour to year timescales.  The one possible short
duration flare points to a duty cycle of less than a few percent if
the flare distribution is uniform.  Alternatively, it is possible that
we observed \lsr\ in a period of relative quiescence, and future
observations may reveal the existence of flares with a higher duty
cycle.

Since our radio observation covers nearly 4 rotation periods of \lsr,
the lack of significant variability points to emission from a uniform
and stable magnetic field configuration.  Assuming a structure with
$R\sim R_*\approx 7\times 10^9$ cm we infer a brightness temperature,
$T_b=2\times 10^9\,F_{\rm \nu,mJy}\,\nu_{\rm GHz}^{-2}\,d_{\rm pc}^2\,
(R/R_*)^{-2}\approx 4\times 10^8$ K, indicative of non-thermal
gyrosynchrotron emission.  This conclusion is supported by the overall
persistence of the radio emission and the low fraction of circular
polarization.

In the context of gyrosynchrotron radiation the emission spectrum is
determined by the size of the emitting region ($R$), the density of
radiating electrons ($n_e$), and the magnetic field strength ($B$)
according to \citep{dm82}:
\begin{equation} 
r_c=0.3\times 10^{1.93{\rm cos}\theta-1.16{\rm cos}^2\theta} 
(3\times 10^3/B)^{-0.21-0.37{\rm sin}\theta},
\end{equation} 
\begin{equation} 
\nu_m=1.8\times 10^4\,({\rm sin}\theta)^{0.5}\,(n_eR)^{0.23} 
\,B^{0.77}\,\,\,{\rm Hz},
\end{equation} 
\begin{equation} 
F_{\nu,m}=2.5\times 10^{-41}\,B^{2.48}\,R^3\,n_e\,({\rm 
sin}\theta)^{-1.52}\,\,\,{\rm \mu Jy}, 
\end{equation} 
where $\theta$ is the angle between the magnetic field and the line of
sight.  Using a range of $\theta=20-70^\circ$ we find $B\lesssim
0.1-20$ G, respectively.  Assuming $\nu_m=8.5$ GHz, we find $R\lesssim
(0.1-7)\times 10^9$ cm, and $n_e\gtrsim 2.4\times 10^{10}-2.3\times
10^{20}$ cm$^{-3}$; the latter is for the range $\theta=70-20^\circ$.
Typical coronal densities in M dwarfs are $\sim 10^{10}-10^{13}$
cm$^{-3}$ \citep{vrm+03,oha+06,bgg+07}, suggesting that $\theta\sim
\pi/3$, and hence $B\sim 10$ G and $R\sim 7\times 10^9\,{\rm cm}\sim
R_*$.  Since the corona is likely to have an overall shell geometry,
the derived physical properties can be interpreted as a structure
extending $\sim 0.3R_*$ above the stellar photosphere with order unity
filling factor.  Smaller filling factors will lead to more extended
magnetic structures.

\begin{figure}
\centerline{\psfig{file=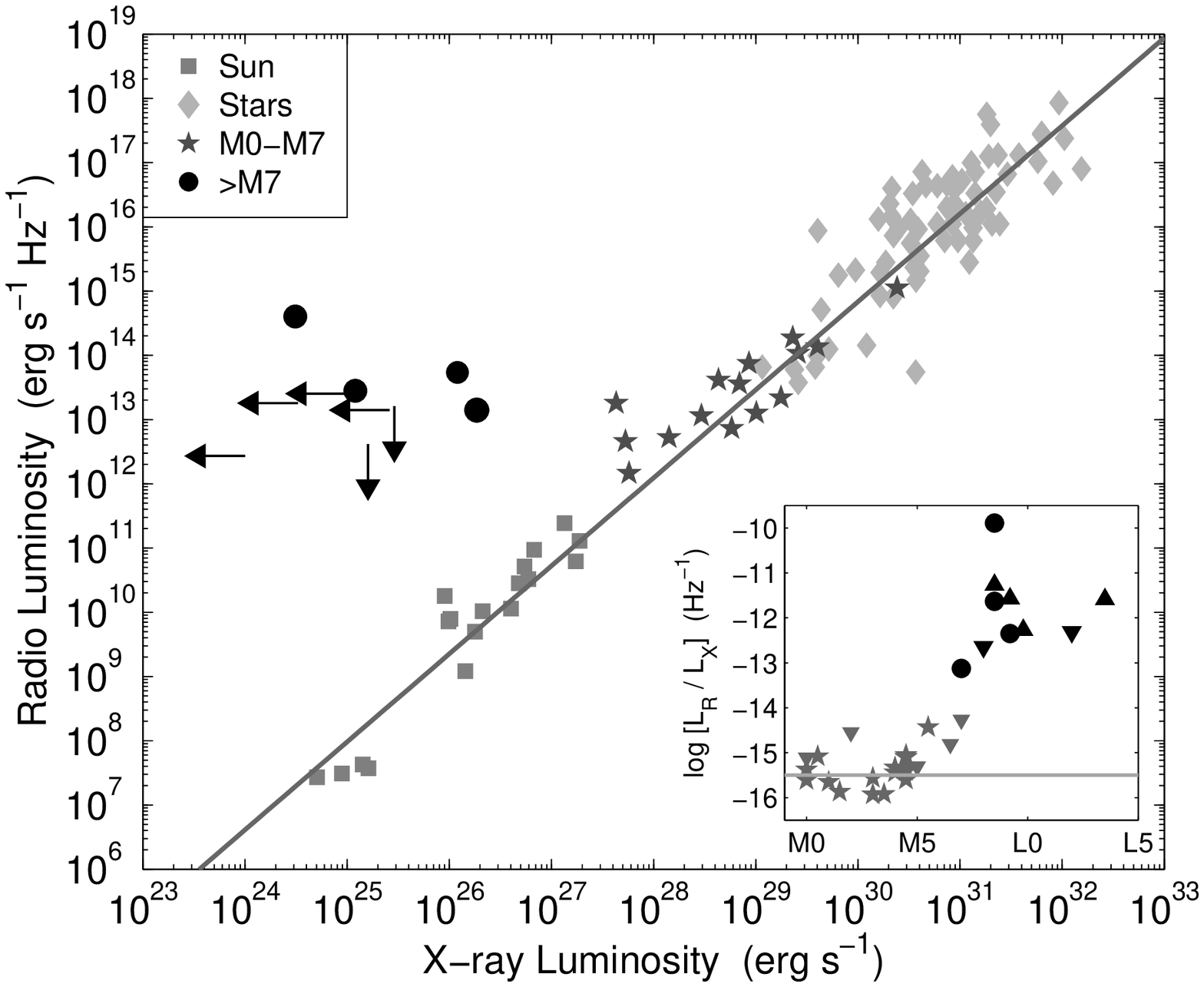,width=6.0in}}
\caption{Radio versus X-ray luminosity for stars exhibiting coronal
activity.  Data for late M and L dwarfs are from \citet{rbm+00},
\citet{bbb+01}, \citet{ber02}, \citet{brr+05}, \citet{bp05},
\citet{ber06}, \citet{aob+07}, and Paper I, while data for other stars
and the Sun are taken from \citet{gud02} and references therein.
Solar data include impulsive and gradual flares, as well as
microflares.  The strong correlation between $L_R$ and $L_X$ is
evident, but it breaks down around spectral type M7 (see inset).  
Of our three targets, \tvlm\ and \lsr\ clearly violate the 
correlation by about four  orders of magnitude, while for \vb\
the upper limit on excess radio emission is less than three
orders of magnitude.
\label{fig:gb}} 
\end{figure}

With strong persistent radio emission and no corresponding X-ray
emission, \lsr\ joins the sample of ultracool dwarfs which strongly
violate the radio/X-ray correlation, $L_R/L_X\approx 10^{-15.5}$
Hz$^{-1}$ \citep{gb93}, observed in F--M6 active stars and in solar
flares.  We find $L_R/L_X\gtrsim 10^{-11.3}$ Hz$^{-1}$, more than four
orders of magnitude in excess of the expected value;
Figure~\ref{fig:gb}.  We stress that the long-term persistence and low
circular polarization of the radio emission clearly point to
gyrosynchrotron radiation from a stellar-scale magnetic field, so the
severe violation cannot be explained away with small-scale,
short-duration heating by coherent processes such as electron
cyclotron maser or plasma radiation.

We next turn to the H$\alpha$ emission for which we measure a range of
equivalent widths of $2-7$ \AA\ (Figure~\ref{fig:lsrall}).  This range
corresponds\footnotemark\footnotetext{To convert from equivalent width
to $L_{\rm H\alpha}/L_{\rm bol}$ we use a conversion factor ($\chi$)
of $10^{-5.3}$ appropriate for \lsr\ \citep{whw04}.} to $L_{\rm
H\alpha}/L_{\rm bol}\approx (1-3.5)\times 10^{-5}$, about an order of
magnitude lower than the saturated H$\alpha$ emission found in mid M
dwarfs.  The weaker chromospheric emission from \lsr\ is in good
agreement with the general trend of decreasing H$\alpha$ emission in
ultracool dwarfs.

As in the case of the radio emission, the level of detected H$\alpha$
emission is puzzling in the context of the non-detected X-ray
emission.  Observations of M0--M6 dwarfs reveal a typical ratio,
$\langle F_{\rm H\alpha}/F_X\rangle\sim 1/3$, with a full range of
$\sim 0.1-1$ \citep{hgr96}.  For \lsr\ this ratio is $F_{\rm H\alpha}/
F_X\gtrsim 5$ (using the weakest H$\alpha$ flux), at least an order of
magnitude larger than in M0--M6 dwarfs.  The significantly brighter
H$\alpha$ emission indicates that chromospheric heating is
significantly more efficient in \lsr\ than coronal heating, possibly
as a result of the field configuration.  We return to this point in
\S\ref{sec:rot}.  Moreover, the elevated H$\alpha$ emission rules out
coronal X-ray emission as the source of chromospheric heating, as
proposed for dMe stars.

While both radio and H$\alpha$ emission are detected, we find a clear
lack of temporal correlation between these two activity indicators.
The H$\alpha$ light curve exhibits significant variability on
timescales of $\sim 0.5-2$ hr, with a wide range of amplitudes,
$\delta {\rm EW}\sim 0.5-5$ \AA.  However, no clear radio variability
is observed in coincidence with any of the H$\alpha$ brightening
episodes.  In particular, during the large increase in H$\alpha$
equivalent width centered on 13:40 UT (Figure~\ref{fig:lsrall}), we
find a limit of $\lesssim 8\%$ on the variability of the radio
emission.  This can be explained in two ways.  First, the H$\alpha$
variability originates from much smaller physical scales than the
radio emission (for which $R\sim R_*$), and any associated radio
variability is therefore dwarfed by the overall persistent flux.  In
this case, we infer that the typical scale of chromospheric regions
giving rise to H$\alpha$ variability is $\lesssim 10\%$ of the overall
chromospheric volume.  Alternatively, the radio emission associated
with the variable H$\alpha$ emission is produced at a much different
frequency than our observed 8.5 GHz band.  This may be the result of
coherent radio emission from fields weaker than $\sim 3$ kG, which
would lead to emission at $\nu=2.8\times 10^6\,B\lesssim 8.5\times
10^9$ Hz.

To summarize, \lsr\ exhibits quiescent radio emission with no
appreciable variability over the timescale of our observation (4
rotation periods), or on year timescales.  The chromospheric H$\alpha$
emission, on the other hand, is highly variable with changes of a
factor of 3 in equivalent width on timescales of $\lesssim 1$ hr.  The
H$\alpha$ flux exceeds the undetected X-ray emission by at least an
order of magnitude, indicating that the chromosphere is not heated by
coronal emission.  The radio emission violates the radio/X-ray
correlation of F--M6 stars by about 4 order of magnitude.

\subsection{\vb} 
\label{sec:vb}

For \vb\ we were able to obtain simultaneous observations only in the
X-rays and UV (Figure~\ref{fig:vb10all}).  The X-ray light curve
exhibits both flaring and quiescent emission with ratios of
$L_X/L_{\rm bol}\approx 10^{-4.1}$ and $10^{-5.0}$, respectively.  As
noted in \S\ref{sec:obs}, quiescent X-ray emission was previously
detected from \vb\ in a shorter observation \citep{fgg03} and we
confirm its existence here at the same flux level.  The X-ray spectrum
is dominated by a $T\approx 3.5\times 10^{6}$ K plasma, with a
possible second component with $T\sim 1.5\times 10^7$ K
(Figure~\ref{fig:vb10xspec}).  These temperatures are typical of
coronal X-ray emission from M dwarfs, and similar to what we
previously found for \tvlm, $T\approx 10^7$ K (Paper I).

The UV emission clearly tracks the X-ray behavior, with detections of
the two flares and the quiescent component.  The peak UV brightness
exceeds the quiescent emission by a factor of about 5
(Table~\ref{tab:vb10uvot}), somewhat less that the order of magnitude
change in X-ray brightness (\S\ref{sec:xrays}).  Moreover, the
quiescent component exceeds the expected photospheric emission by
about an order of magnitude, and thus points to the existence of a
persistent transition region, as noted previously by \citet{hj03}.  So
far this is the only case of correlated emission found in any of our
targets, suggesting that at least some late M dwarfs may follow the
behavior seen in early M dwarfs.

As in the case of the X-ray and UV data, our non-simultaneous optical
spectroscopy also reveals a significant level of variability.  As far
as we know, our spectroscopic observations provide the most extensive
coverage of \vb\ in the published literature, and are thus valuable
even if no multi-wavelength coverage is available.  Here we provide a
brief summary of the observations, and defer a detailed analysis to a
future publication.

We clearly detect H$\alpha$ in emission in all of the individual
exposures.  An equivalent width light curve is shown in
Figure~\ref{fig:vb10balmer}.  The range of measured values is
$3.7-8.6$ \AA, extending to somewhat higher values than previously
published for \vb.  Similarly, the H$\beta$ line varies from 5 to 26
\AA, while H$\gamma$ and H$\delta$ range from undetectable levels
($\lesssim {\rm few}$ \AA) to about 80 \AA.  We also detect
\ion{Ca}{2} H\&K emission ranging from non-detectable levels to a K
line\footnotemark\footnotetext{The \ion{Ca}{2} H line is blended with
H$\epsilon$.} equivalent width of about 30 \AA.

The emission line light curves exhibit a wide range of variability
timescales.  We note that the total duration of our observation
samples only about 1/4 of a rotation of \vb\ ($v{\rm sin}i=6.5$ km
s$^{-1}$), and we therefore have no information on rotational
modulation, which was detected in our previous target, \tvlm\ (Paper
I).  The variability is primarily gradual with the rise and fall times
of individual episodes being roughly the same.  The broadest
brightening has a rise time of about 0.7 hr and a decay time of about
2 hr, with a factor of two increase in equivalent width.  The
variability is reflected in all of the detected Balmer lines, but the
relative change in equivalent width is largest for the weakest lines.
This mild change in the Balmer decrement points to increased densities
and temperatures during brighter emission episodes.  Sample spectra in
these High and Low states are shown in Figure~\ref{fig:vb10spec}.

In addition to the time-resolved and gradual changes in emission line
equivalent width we detect two episodes of rapid brightening with a
duration of $\lesssim 300$ s (Flare state).  These flares are marked
by a large increase in the equivalent width of the higher-order Balmer
lines and the \ion{Ca}{2} lines relative to the High state spectra.
They are further distinguished from the High state spectra by the
appearance of \ion{He}{1} lines (Figure~\ref{fig:vb10spec}).
Temporally, both of the detected flares are superposed on gradual
brightenings, suggesting that the two modes of variability are
causally and temporally unrelated and arise in different regions of
the chromosphere.

The significant difference in Balmer decrement between the High and
Flare states, and the appearance of \ion{He}{1} lines in the latter
indicate higher plasma densities, and most likely temperatures in the
Flare state.  A comparison to flares on mid M dwarfs (\citealt{aha+06}
and references therein) reveals shallower Balmer decrements in the
impulsive flares on \vb: ${\rm H\delta}/{\rm H\gamma}\approx 0.9$,
${\rm H\epsilon}/{\rm H\gamma} \approx 0.9$, and ${\rm H\xi}/{\rm
H\gamma}\approx 0.6$, compared to average values of 0.8, 0.5, and 0.4,
respectively.  Similarly, the formation of \ion{He}{1} lines generally
requires dense plasma ($n_e\sim 10^{13}-10^{16}$ cm$^{-3}$) at $T\sim
{\rm few}\times 10^4$ K, approaching the transition region
temperatures.

Although our X-ray and optical observations are not simultaneous, the
variability timescales in both bands are similar
(Figures~\ref{fig:vb10all} and \ref{fig:vb10balmer}).  This suggests
that in \vb\ chromospheric heating by coronal X-rays may in fact work.
Indeed, the observed range of $L_{\rm H\alpha}/L_{\rm bol}\approx
10^{-4.5}-10^{-4.9}$ (Table~\ref{tab:vb10spec}) is similar to the
range of $L_X/L_{\rm bol}\approx 10^{-4.1}-10^{-5.0}$.  Thus, the
chromospheric and coronal structure of \vb\ is markedly different than
that of \lsr, for which the H$\alpha$ luminosity exceeds the X-ray
luminosity by at least an order of magnitude, and no evidence for
plasma at coronal temperatures exists.

To conclude, the magnetic activity in \vb\ differs significantly from
that of \lsr\ and \tvlm, and is more typical of M0--M6 dwarfs.  The
X-ray and UV emission are clearly correlated and point to the presence
of quiescent and flaring corona and transition region.  The optical
chromospheric emission exhibits similar timescales and amplitudes of
variability, and can be separated into a gradually variable component
(Low and High states), and a rapidly flaring component (Flare state)
that is marked by higher plasma densities and temperatures.

\section{The Possible Role of Rotation}
\label{sec:rot}

With the increasing sample of ultracool dwarfs observed in the radio
and X-rays we can begin to address the physical conditions that
underlie the generation of magnetic fields and their influence on the
coronae and chromospheres.  It is clear from our work that late M
dwarfs exhibit a mix of activity patterns in comparison to stars with
spectral types F to early M stars, whose magnetic activity is powered
by the $\alpha\Omega$ dynamo.  Thus, it is instructive to investigate
whether rotation also plays a role in the magnetic activity of
ultracool dwarfs.

\subsection{X-ray Activity}
\label{sec:xr}

\begin{figure}
\centerline{\psfig{file=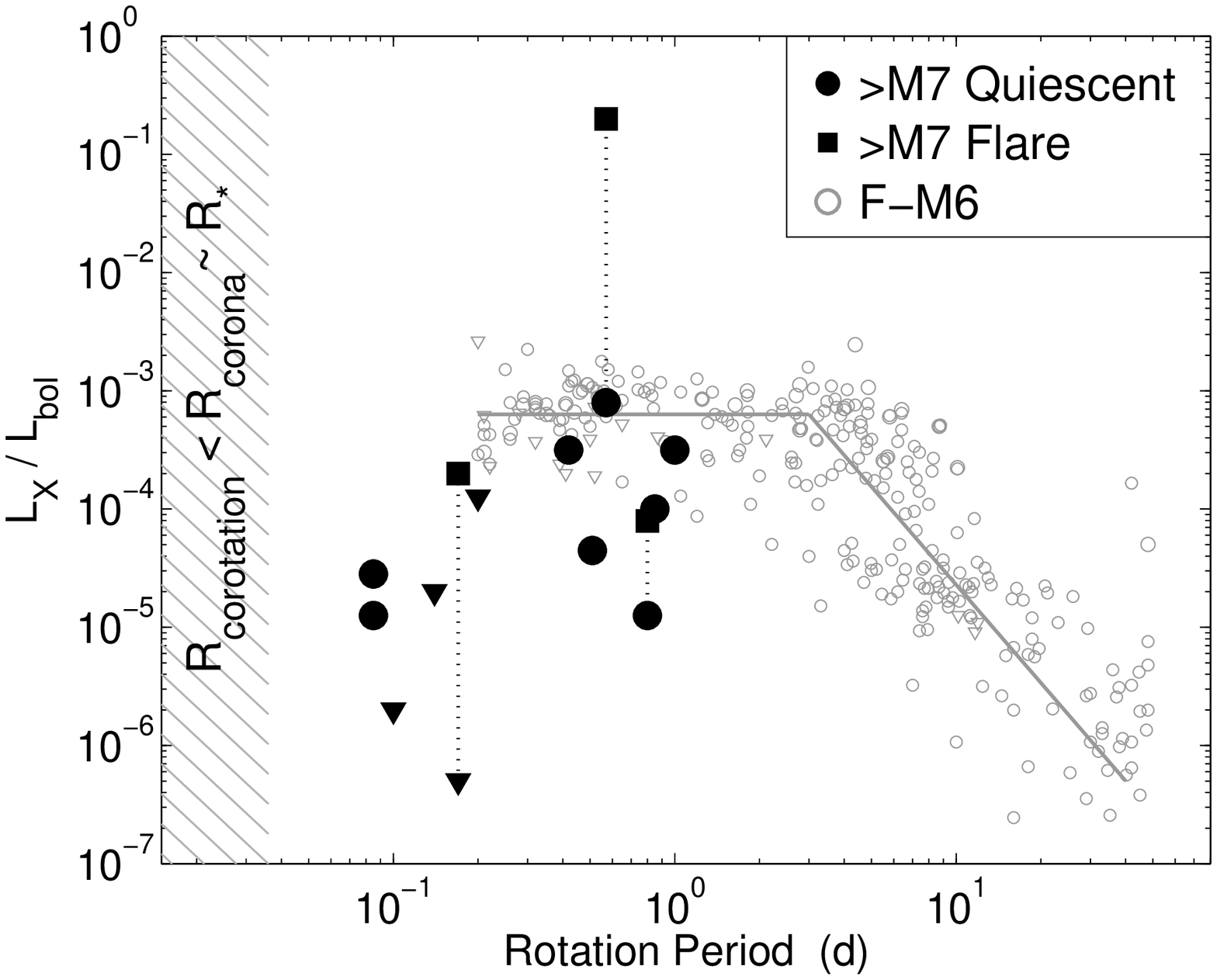,width=6.0in}}
\caption{Ratio of X-ray to bolometric luminosity as a function of
rotation period for stars of spectral type F--M6 (gray circles), and
$>{\rm M7}$ (black symbols).  Data are from \citet{jjj+00},
\citet{pmm+03}, \citet{rb07}, \citet{aob+07} and references therein,
and this paper.  The rotation-activity relation and saturation in the
F--M6 stars are clearly seen (gray lines).  The X-ray activity of 
ultracool dwarfs is generally weaker than in early type stars, and in 
addition appears to {\it decrease} at higher rotation velocity.  
This serves as possible evidence for super-saturation.  The shaded 
region marks the rotation period below which the corotation radius is 
smaller than the coronal radius (assumed to extend $\sim R_*$ above 
the stellar surface), leading to possible stripping of coronal 
material by centrifugal ejection.  
\label{fig:lxrot}}
\end{figure}

The generation of persistent X-ray emission in F--M stars is still not
fully understood, but its correlation with radio activity and rotation
velocity suggests that it is directly tied to heating by magnetic
field dissipation and particle acceleration.  In
Figure~\ref{fig:lxrot} we show the well-known rotation-activity
relation and saturation of $L_X/L_{\rm bol}$ for stars with spectral
type F--M6.  The rotation-activity relation provides evidence for the
existence of an $\alpha\Omega$ dynamo, while three primary
explanations have been proposed for the saturation of $L_X/L_{\rm
bol}$ (see also \citealt{gud04}).  The first is saturation of the
dynamo mechanism itself at low Rossby numbers, ${\rm Ro}\equiv
P/\tau_c\lesssim 0.1$ (e.g., \citealt{vil84}); here $\tau_c$ is the
convective turnover time.  Second, centrifugal effects in rapid
rotators may lead to a circulation pattern that sweeps the magnetic
fields from the lower convection zone toward the poles, thereby
reducing the field filling factor \citep{ssv01}.

The third explanation is centrifugal stripping of the corona at the
corotation radius\footnotemark\footnotetext{The corotation radius,
$R_c=(GM_*/\Omega^2)^{1/3}$, is the radius at which the centrifugal
and gravitational forces are exactly balanced.}, $R_c$
\citep{ju99,jjj+00}.  In this scenario, the particle density,
$n_e\propto\Omega$, and emission volume, $V\propto R_c^3\propto
\Omega^{-2}$, conspire to produce an emission measure ($\propto
n_e^2V$) that is independent of the rotation period, i.e., saturation.
However, this balance breaks down when the corotation radius is well
inside the corona, $R_c\lesssim R_*$, leading to super-saturation -- a
reduction in $L_X/L_{\rm bol}$ at very rapid rotation rates (e.g.,
\citealt{jjj+00}).

It is instructive to study whether the X-ray emission of ultracool
dwarfs exhibits saturation and/or super-saturation, and if so, which
of the mechanisms described above is responsible for these effects.
The values of $L_X/L_{\rm bol}$ as a function of rotation period
presented here and in Paper I, as well as in the literature, are
summarized in Figure~\ref{fig:lxrot}.  Two trends are clear.  In the
range of rotation periods of $\sim 0.4-1$ d, in which F--M6 stars
exhibit saturated emission, $L_X/L_{\rm bol}\sim 10^{-4}$ for the
ultracool dwarfs is a few times smaller than the saturated value of
the F--M6 stars.  This points to generally weaker coronae in ultracool
dwarfs compared to higher mass stars, although in one case an
unusually bright flare with $L_X/L_{\rm bol}\sim 0.1$ was detected
\citep{ssm+06}.  On the other hand, ultracool dwarfs with periods of
$\lesssim 0.2$ d ($v{\rm sin}i\gtrsim 25$ km s$^{-1}$) exhibit much
lower values of $L_X/L_{\rm bol}\lesssim 10^{-5}$.  These include our
targets \lsr\ and \tvlm.  The reduction in X-ray activity at high
rotation velocities is reminiscent of the super-saturation phenomenon.

In the context of centrifugal stripping, an ultracool dwarf with
$M\sim 0.1$ M$_\odot$ and $R\sim 0.1$ R$_\odot$ has a corotation
radius that is smaller than the coronal radius ($\sim R_*$ above the
photosphere) when $P\lesssim 1$ hr.  This is a factor of two times
smaller than the fastest rotators in the sample, with $v{\rm
sin}i\approx 60$ km s$^{-1}$.  This indicates that centrifugal
stripping may play a role in reducing $L_X/L_{\rm bol}$ for fast
rotators only if the the magnetic field is generally dominated by
extended loops with $R\sim (2-3)R_*$, which may indeed be the relevant
scale for M dwarfs (e.g., \citealt{lpl+00}).  In the case of \lsr,
\tvlm\ (Paper I), and 2MASS\,J00361617+1821104 \citep{brr+05} we
inferred from our radio observations a magnetic field scale of $\sim
(1-{\rm few})\times 10^{10}$ cm, in the rough range required for
efficient centrifugal stripping.

Thus, to the extent that existing observations of ultracool dwarfs
reveal possible super-saturation in the X-rays, it is presently
unclear whether this is due to intrinsic dynamo effects or to the
secondary effect of centrifugal stripping.  The proposed mechanism of
centrifugal clearing of the equatorial plane in rapid rotators
\citep{ssv01} is not likely to play a role since the persistent radio
emission on $\gtrsim {\rm year}$ timescales in several ultracool
dwarfs points to fields with order unity covering fractions.

An alternative possibility for the drop in $L_X/L_{\rm bol}$ is that
the source of coronal heating is different, and much less efficient,
than in higher mass stars.  This scenario may explain the lower values
of $L_X/L_{\rm bol}$ in the regime where F--M6 stars are saturated.
However, it is unclear why such a scenario should depend on rotation,
particularly since the radio emission (i.e., particle acceleration)
does not diminish in fast rotators (see \S\ref{sec:radio}).

We caution that the possibility of super-saturation in the X-rays is
based on a small number of objects, and moreover that the effect of
centrifugal stripping depends on the magnetic field configuration.
Future X-ray observations of additional late M and L dwarfs, as well
as the determination of rotation velocities for a larger sample of
ultracool dwarfs will reveal whether, and at what rotation velocity,
super-saturation may set in present.  If confirmed, the value of
$v{\rm sin}i$ at which super-saturation occurs may hold a clue to the
nature of the dynamo or the dominant physical scale of coronal
magnetic loops.

\subsection{Radio Activity}
\label{sec:radio}

\begin{figure}
\includegraphics[angle=90,width=6.8in]{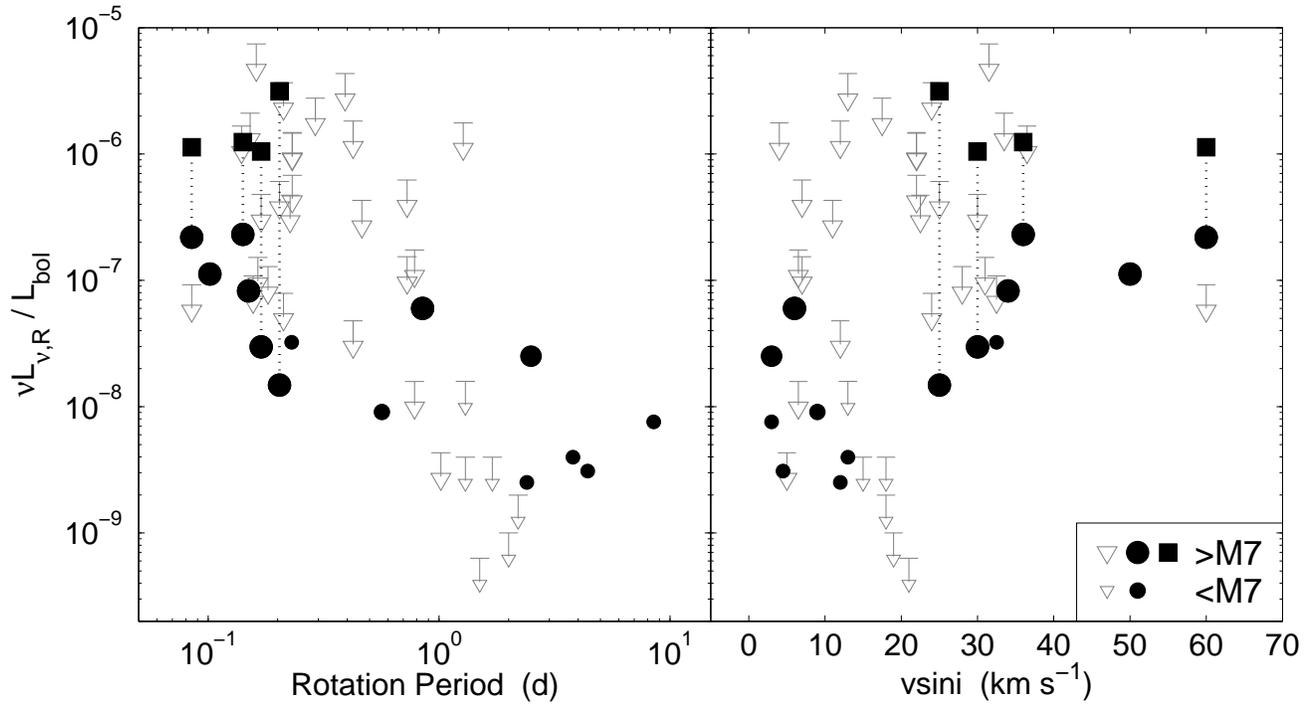}
\caption{Ratio of radio to bolometric luminosity as a function of
rotation period ({\it Left}) and rotation velocity ({\it Right}).  
Data are from \citet{wjk89}, \citet{ber02}, \citet{ber06} and 
references therein, and this paper.  There is an apparent 
correlation between radio activity and rotation, with no clear 
sign of saturation or super-saturation.
\label{fig:lrrot}}
\end{figure}

Since radio emission is arguably a more robust tracer of magnetic
activity than X-ray emission, we repeat the same exercise for
$L_R/L_{\rm bol}$ as a function of rotation period and velocity.
\citet{ber02} noted a possible correlation between radio activity and
rotation based on a small sample of ultracool dwarfs, and we re-visit
this investigation here.  In Figure~\ref{fig:lrrot} we plot radio
detections and upper limits for objects with a known rotation velocity
and period.  We find that for objects with spectral type $>{\rm M7}$
those with $v{\rm sin}i\lesssim 30$ km s$^{-1}$, or $P\gtrsim 4$ hr,
exhibit a typical quiescent radio activity, $\nu L_{\rm \nu ,R}/L_{\rm
bol}\sim 10^{-7.5}$.  On the other hand, those with $v{\rm
sin}i\gtrsim 30$ km s$^{-1}$ have $\nu L_{\rm \nu ,R}/L_{\rm bol}\sim
10^{-6.8}$, nearly an order of magnitude larger.  If we include
objects earlier than M7, the increase in radio activity with rotation
velocity becomes even more pronounced, with a typical level of $\nu
L_{\rm \nu ,R}/L_{\rm bol}\lesssim 10^{-8}$ for $v{\rm sin}i\lesssim
30$ km s$^{-1}$.

The majority of the ultracool dwarfs with a known rotation velocity
remain undetected in VLA observations with a typical duration of $\sim
2$ hr, mostly as a result of the decrease in $L_{\rm bol}$ for L and T
dwarfs (Figure~\ref{fig:lrrot} and \citealt{ber06}).  Still, it
appears that with the present sample, there is possible evidence for a
radio rotation-activity relation, with no evidence for either
saturation or super-saturation.  This is contrary to the apparent
decrease in $L_X/L_{\rm bol}$ at $v{\rm sin}i\sim 25$ km s$^{-1}$
(\S\ref{sec:xr}).  If confirmed with future observations, the most
likely explanation is that the dynamo mechanism is at least partially
driven by rotation (even in the presence of full convection), with no
apparent saturation at very low Rossby numbers (${\rm Ro}\lesssim
10^{-3}$), contrary to the trend in F--M6 stars.

Regardless of the exact dynamo mechanism there is a clear discrepancy
between the radio and X-ray observations in terms of a possible
activity-rotation relation and its super-saturation.  The lack of
clear saturation or super-saturation in the radio indicates that the
dynamo itself is not likely to saturate even at high rotation
velocities.  The possible decline in X-ray activity at $v{\rm
sin}i\gtrsim 25$ km s$^{-1}$ may thus be the result of secondary
effects such as inefficient heating of the plasma to coronal
temperatures, or centrifugal stripping of the most extended magnetic
loops, with $R\sim {\rm few}\times R_*$.  The former scenario is
supported by the observed ratio $L_{\rm H\alpha}/L_X\gtrsim 1$ in the
quiescent emission from \lsr\ and \tvlm, which points to heating of
chromospheric plasma by a process other than coronal X-ray emission.

The difference in trends between radio and X-ray activity as a
function of rotation velocity may also underlie the violation of the
radio/X-ray correlation in ultracool dwarfs (Figure~\ref{fig:gb}).  In
Figure~\ref{fig:bgrot} we plot $L_{\nu,R}/L_X$ relative to its
standard value in F--M6 stars ($\approx 10^{-15.5}$; \citealt{gb93})
as a function of rotation velocity.  We find that objects with $v{\rm
sin}i\gtrsim 30$ km s$^{-1}$ exhibit a significantly more severe
violation of the radio/X-ray correlation than those with slower
rotation.  It is important to note, however, that rotation alone may
not fully account for the excess ratios since even objects with $v{\rm
sin}i\lesssim 10$ km s$^{-1}$ may exceed the value of $10^{-15.5}$ by
an order of magnitude or more.  Still, it appears that faster rotators
generally violate the radio/X-ray correlation by a larger factor.

\section{Discussion and Summary}
\label{sec:conc}

We presented simultaneous multi-wavelength observations of two late M
dwarfs to trace the magnetic activity in their outer atmospheres.  In
the case of \lsr\ (M8.5) we detect persistent radio emission and
highly variable H$\alpha$ emission.  No excess UV emission is
detected.  Similarly, we detect no X-ray emission to a deep limit of
$L_X/L_{\rm bol}\lesssim 10^{-5.7}$.  The ratio of radio to X-ray flux
exceeds the average value measured in a wide range of active stars by
more than 4 orders of magnitude.  Similarly, the ratio $L_{\rm
H\alpha}/L_X\gtrsim 10$ exceeds the values measured for M0--M6 dwarfs
by at least an order of magnitude, and rules out heating of the
quiescent chromosphere by coronal emission.

Temporally, we find no correspondence between the radio and H$\alpha$
light curves.  This indicates that even if the source of chromospheric
heating is magnetic reconnection, it occurs on sufficiently small
scales that the overall radio emission does not change by more than
$\sim 10\%$.  Alternatively, the radio emission produced in
conjunction with the H$\alpha$ variability may peak at a lower
frequency than our 8.5 GHz observations.  In the former scenario, our
limit on the radio variability during the brightest H$\alpha$ episode
can be interpreted as a limit of $\lesssim 10\%$ on the chromospheric
volume involved in the variable H$\alpha$ emission.  In the latter
scenario, it is possible that the coincident radio emission is
dominated by coherent emission at $\nu\approx 2.8\times 10^6\,B$, with
$B\lesssim 3$ kG.

With our measured rotation velocity of $v{\rm sin}i\approx 50$ km
s$^{-1}$ for \lsr, the radio observations sample nearly 4 full
rotations.  The stability of the radio emission thus requires a
large-scale and uniform magnetic field, with an inferred strength of
$\sim 10$ G.  The similarity in flux to a past observation suggests
that the field is stable on year timescales.  This may be expected
given the long convective turnover time for ultracool dwarfs, on the
order of several years.

\vb, on the other hand, exhibits bright and variable X-ray and UV
emission, including a pair of flares and quiescent emission.  The
behavior in both bands, which trace the corona and transition region,
respectively, is highly correlated.  From the X-ray spectrum we infer
a coronal temperature of about $3\times 10^6$ K, with a possible
second component at $\sim 10^7$ K.  The optical emission lines exhibit
a similar amplitude and timescale of variability to the X-rays and UV,
with two distinct states of gradual and impulsive flaring.  The lack
of causal relation between the two states points to emission from
distinct plasma environments.  Moreover, the shallow Balmer decrement
and \ion{He}{1} emission in the impulsive Flare state require
significantly denser and hotter plasma.

Our detailed study of \lsr, \vb, and \tvlm\ reveals mixed patterns of
of behavior compared to the magnetic activity in F-M6 stars, and thus
points to a transition in the atmospheric structure and heating
process in the late M dwarf regime.  In particular, stellar-scale
magnetic fields are present, as evidence by quiescent and uniform
radio emission, but the coronae are generally weaker than in early M
dwarfs.  The chromospheric activity declines less rapidly, and all
three targets exhibit a similar range of highly variable H$\alpha$
emission, $L_{\rm H\alpha}/L_{\rm bol}\sim 10^{-5}-10^{-4.5}$.  This
range is about an order of magnitude less that the saturated value of
mid M dwarfs.

In the standard picture of solar and stellar magnetic flares, the
release of magnetic energy through processes such as reconnection
leads to acceleration of electrons, and subsequently heating of the
chromosphere, transition region, and corona through evaporation of the
lower atmosphere.  In quiescence, the chromospheres of M0--M6 dwarfs
may instead be heated by coronal X-ray emission \citep{cra82}, as
evidenced by the typical observed ratios of $L_{\rm H\alpha}/L_X\sim
1/3$ \citep{cra82,hgr96}.  Such a mechanism is clearly not at play in
the case of \lsr, and most likely \tvlm, since the chromospheric
emission is significantly more energetic than the non-detected corona.

It is possible instead that the source of chromospheric heating is
continuous micro-flaring activity, leading to replenishment of
chromospheric and transition region plasma by evaporation.  In this
scenario the weak or absent coronal emission may be due to a limited
temperature enhancement of $\lesssim 10^5$ K, which is sufficient to
produce chromospheric emission, but no significant soft X-ray
emission.  The continuously variable H$\alpha$ emission in \lsr\
appears to support this idea, and along with the lack of corresponding
radio variability points to heating on scales much smaller than the
stellar photosphere.

\begin{figure}
\centerline{\psfig{file=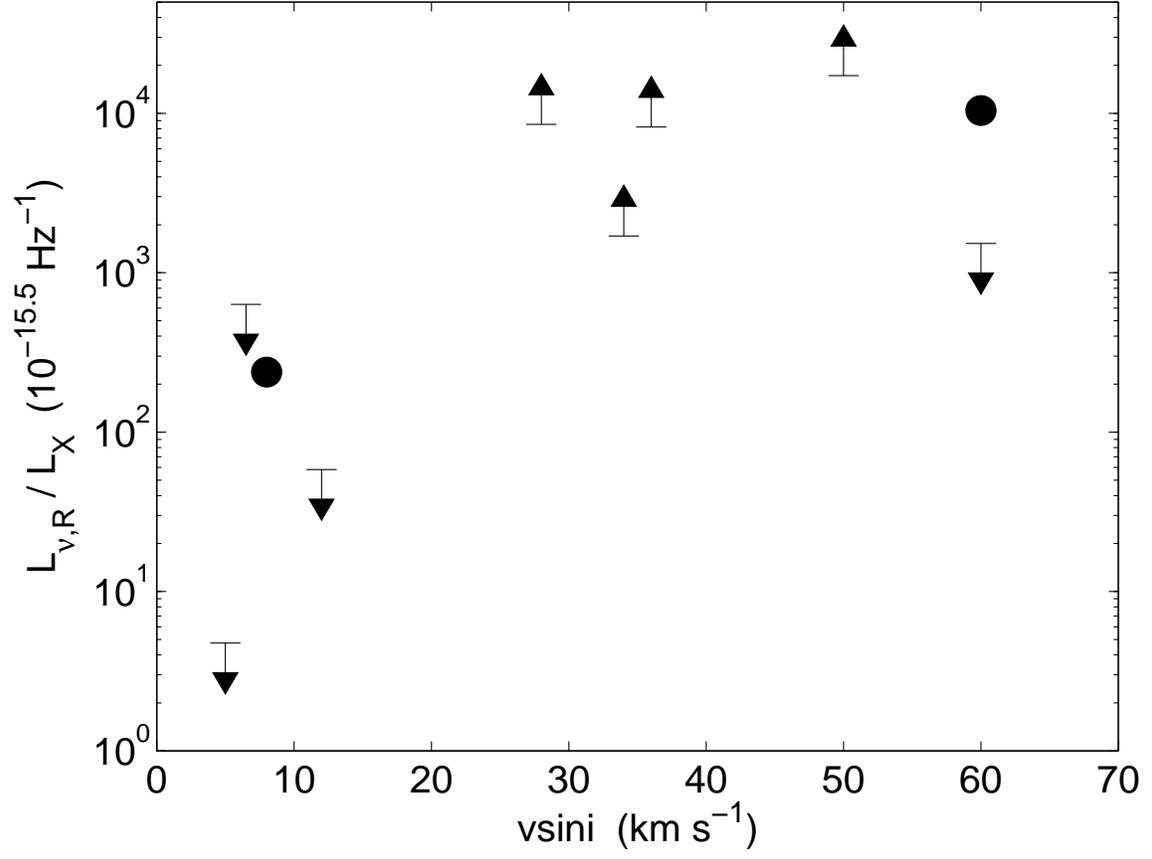,width=6.0in}}
\caption{Ratio of radio to X-ray luminosity normalized to the typical
value observed in F--M6 stars ($L_{\rm \nu,R}/L_X\approx 10^{-15.5}$;
Figure~\ref{fig:gb}) as a function of rotation velocity.  Shown are
objects with spectral type $>{\rm M7}$.  It appears that the most
severe violators of the radio/X-ray correlation are those with the
highest rotation velocity.
\label{fig:bgrot}} 
\end{figure}

As we noted in \S\ref{sec:rot}, it is also possible that rapid
rotation in ultracool dwarfs suppresses the X-ray emission through
centrifugal stripping.  The apparent increase in radio activity with
rotation indicates that the dynamo itself does not saturate, at least
up to $v{\rm sin}i\sim 60$ km s$^{-1}$, which is roughly $1/3$ of the
break-up velocity.  The effects leading to the apparent
super-saturation in the X-rays may also underlie the severe violation
of the radio/X-ray correlation since the level of violation appears to
be correlated with rotation velocity (Figure~\ref{fig:bgrot}).  We
stress that the role of rotation is still speculative due to the small
number of objects with X-ray, radio, and rotation measurements.  It is
therefore crucial to increase the sample of ultracool dwarfs for which
these quantities are measured.

To summarize, we conclude that late M dwarfs mark a change in the
properties of the magnetic field and its dissipation, as well as the
generation of high temperature plasma in the outer atmosphere.  The
weak X-ray emission, both in relation to $L_{\rm bol}$ and $L_{\rm
H\alpha}$, points to inefficient heating of plasma to coronal
temperatures, or alternatively to stripping of the most extended
magnetic loops.  The lack of temporal correlation between the radio
and H$\alpha$ activity, however, indicates that even if atmospheric
evaporation is taking place, it occurs on smaller physical scale than
the overall magnetic field structure.

We end by noting that the use of simultaneous, multi-wavelength, and
long duration observations to probe the magnetic activity of
individual ultracool dwarfs provides a crucial complement to large
single-band surveys.  In particular, the long duration and high time
resolution data elucidate the range of timescales and amplitudes of
gradual and impulsive flares, and can moreover uncover
rotationally-induced modulations, as in the case of \tvlm\ (Paper I).
In addition, the existence or absence of correlations between the
various activity indicators (as in \lsr\ and \tvlm), and their
implications for the magnetic field and its dissipation, could not
have been inferred from existing observations.  In the upcoming
Chandra cycle we will expand our analysis with observations of several
L0--L3 dwarfs.  These observations will reveal whether the transition
in magnetic activity patterns seen in the late M dwarfs continues to
lower mass objects, some of which are bona-fide brown dwarfs.

\acknowledgements We thank the Chandra, Gemini, VLA, and Swift
schedulers for their invaluable help in coordinating these
observations.  This work has made use of the SIMBAD database, operated
at CDS, Strasbourg, France.  It is based in part on observations
obtained at the Gemini Observatory, which is operated by the
Association of Universities for Research in Astronomy, Inc., under a
cooperative agreement with the NSF on behalf of the Gemini
partnership: the National Science Foundation (United States), the
Science and Technology Facilities Council (United Kingdom), the
National Research Council (Canada), CONICYT (Chile), the Australian
Research Council (Australia), CNPq (Brazil) and CONICET (Argentina).
Data from the UVOT instrument on Swift were used in this work.  Swift
is an international observatory developed and operated in the US, UK
and Italy, and managed by NASA Goddard Space Flight Center with
operations center at Penn State University.  Support for this work was
provided by the National Aeronautics and Space Administration through
Chandra Award Number G07-8014A issued by the Chandra X-ray Observatory
Center, which is operated by the Smithsonian Astrophysical Observatory
for and on behalf of the National Aeronautics Space Administration
under contract NAS8-03060.

\clearpage
\begin{deluxetable}{lll}
\tablecolumns{3}
\tabcolsep0.1in\footnotesize
\tablewidth{0pc}
\tablecaption{UVOT/UVW1 magnitudes of \vb\
\label{tab:vb10uvot}}
\tablehead{
\colhead{Exposure}       &
\colhead{UT Time}        &
\colhead{AB Mag$^a$}     \\
}
\startdata
1 & 11:08:22--11:31:00 & $19.91\pm 0.12$ \\
2 & 12:32:23--12:49:00 & $20.45\pm 0.21$ \\
3 & 14:08:27--14:36:00 & $<21.47$ \\
4 & 15:44:28--16:12:00 & $<21.40$ \\
5 & 17:21:27--17:49:00 & $<21.40$ \\
3--5 & 14:08:27--17:49:00 & $21.57\pm 0.24$ \\
6 & 18:57:27--19:15:00 & $21.07\pm 0.35$
\enddata
\tablecomments{$^a$ Limits are $3\sigma$.}
\end{deluxetable}

\begin{deluxetable}{lccccccccc}
\tablecolumns{10}
\tabcolsep0.1in\footnotesize
\tablewidth{0pc}
\tablecaption{Emission Line Properties of \vb\
\label{tab:vb10spec}}
\tablehead{
\colhead {}        &
\multicolumn{3}{c}{Low} &
\multicolumn{3}{c}{High} &
\multicolumn{3}{c}{Flare} \\\cline{2-4}\cline{5-7}\cline{8-10}
\colhead{Line}            &
\colhead{EW$^a$}          &
\colhead{Flux$^b$}        &
\colhead{${\rm log}(L/L_{\rm bol})$} &
\colhead{EW}              &
\colhead{Flux}            &
\colhead{${\rm log}(L/L_{\rm bol})$} &
\colhead{EW}              &
\colhead{Flux}            &
\colhead{${\rm log}(L/L_{\rm bol})$}
}
\startdata
H$\alpha$                   & 4.3  & 5.2  & $-4.9$ & 8.6  & 13.3 & $-4.5$ & 9.0 & 15.1 & $-4.4$  \\
H$\beta$                    & 4.8  & 0.5  & $-5.9$ & 17   & 2.8  & $-5.1$ & 33  & 8.7  & $-4.6$  \\
H$\gamma$                   & 9.3  & 0.2  & $-6.3$ & 31   & 1.6  & $-5.4$ & 75  & 8.7  & $-4.6$  \\
H$\delta$                   & \nod & \nod & \nod   & 31   & 1.0  & $-5.6$ & 95  & 8.0  & $-4.7$  \\
H$\epsilon$+\ion{Ca}{2}\,H  & 8.5  & 0.2  & $-6.3$ & 111  & 1.8  & $-5.3$ & 83  & 7.8  & $-4.7$  \\
H$\xi$                      & \nod & \nod & \nod   & 9.1  & 0.3  & $-6.1$ & 44  & 4.8  & $-4.9$  \\
\ion{Ca}{2}\,K              & 15   & 0.2  & $-6.3$ & 39   & 1.2  & $-5.5$ & 26  & 2.5  & $-5.2$  \\
\ion{He}{1}\,$\lambda 5877$ & \nod & \nod & \nod   & \nod & \nod & \nod   & 2.8 & 1.1  & $-5.6$  \\
\ion{He}{1}\,$\lambda 4473$ & \nod & \nod & \nod   & \nod & \nod & \nod   & 5.3 & 0.9  & $-5.6$  \\
\ion{He}{1}\,$\lambda 4027$ & \nod & \nod & \nod   & \nod & \nod & \nod   & 5.4 & 0.5  & $-5.9$  
\enddata
\tablecomments{Emission line fluxes, equivalent widths, and ratios
relative to the bolometric luminosity for the three spectral states
identified in \vb\ (Figures~\ref{fig:vb10balmer} and 
\ref{fig:vb10spec}).\\
$^a$ In units of \AA.\\
$^b$ In units of $10^{-15}$ erg cm$^{-2}$ s$^{-1}$ \AA$^{-1}$.}
\end{deluxetable}

\clearpage

\clearpage

\clearpage

\clearpage

\clearpage

\clearpage

\clearpage

\clearpage

\clearpage

\clearpage

\clearpage

\clearpage


\begin{thebibliography}{}

\bibitem[\protect\citeauthoryear{{Allard} et~al.}{{Allard}
  et~al.}{2001}]{aha+01}
{Allard}, F., {Hauschildt}, P.~H., {Alexander}, D.~R., {Tamanai}, A.,  \&
  {Schweitzer}, A. 2001, \apj, 556, 357

\bibitem[\protect\citeauthoryear{{Allred} et~al.}{{Allred}
  et~al.}{2006}]{aha+06}
{Allred}, J.~C., {Hawley}, S.~L., {Abbett}, W.~P.,  \& {Carlsson}, M. 2006,
  \apj, 644, 484

\bibitem[\protect\citeauthoryear{{Antonova} et~al.}{{Antonova}
  et~al.}{2007}]{adh+07}
{Antonova}, A., {Doyle}, J.~G., {Hallinan}, G., {Golden}, A.,  \& {Koen}, C.
  2007, astro-ph/0707.0634

\bibitem[\protect\citeauthoryear{{Aschwanden} et~al.}{{Aschwanden}
  et~al.}{2001}]{apr01}
{Aschwanden}, M.~J., {Poland}, A.~I., \& {Rabin}, D.~M. 2001, \araa, 
39, 175

\bibitem[\protect\citeauthoryear{{Audard} et~al.}{{Audard}
  et~al.}{2007}]{aob+07}
{Audard}, M., {Osten}, R.~A., {Brown}, A., {Briggs}, K.~R., {Guedel}, M.,
  {Hodges-Kluck}, E.,  \& {Gizis}, J.~E. 2007, astro-ph/0707.1882

\bibitem[\protect\citeauthoryear{{Benz} \& {Guedel}}{{Benz} \&
  {Guedel}}{1994}]{bg94}
{Benz}, A.~O.,  \& {Guedel}, M., 285, 621

\bibitem[\protect\citeauthoryear{{Berger}}{{Berger}}{2002}]{ber02}
{Berger}, E. 2002, \apj, 572, 503

\bibitem[\protect\citeauthoryear{{Berger}}{{Berger}}{2006}]{ber06}
{Berger}, E. 2006, \apj, 648, 629

\bibitem[\protect\citeauthoryear{{Berger} et~al.}{{Berger}
  et~al.}{2001}]{bbb+01}
{Berger}, E., et~al. 2001, \nat, 410, 338

\bibitem[\protect\citeauthoryear{{Berger} et~al.}{{Berger}
  et~al.}{2007}]{bgg+07}
{Berger}, E., et~al. 2007, astro-ph/0708.1511 (Paper I)

\bibitem[\protect\citeauthoryear{{Berger} et~al.}{{Berger}
  et~al.}{2005}]{brr+05}
{Berger}, E., et~al. 2005, \apj, 627, 960

\bibitem[\protect\citeauthoryear{{Burgasser} \& {Putman}}{{Burgasser} \&
  {Putman}}{2005}]{bp05}
{Burgasser}, A.~J.,  \& {Putman}, M.~E. 2005, \apj, 626, 486

\bibitem[\protect\citeauthoryear{{Cram}}{{Cram}}{1982}]{cra82}
{Cram}, L.~E. 1982, \apj, 253, 768

\bibitem[\protect\citeauthoryear{{Dulk} \& {Marsh}}{{Dulk} \&
  {Marsh}}{1982}]{dm82}
{Dulk}, G.~A.,  \& {Marsh}, K.~A. 1982, \apj, 259, 350

\bibitem[\protect\citeauthoryear{{Fleming}, {Giampapa}, \& {Garza}}{{Fleming}
  et~al.}{2003}]{fgg03}
{Fleming}, T.~A., {Giampapa}, M.~S.,  \& {Garza}, D. 2003, \apj, 594, 982

\bibitem[\protect\citeauthoryear{{Fleming}, {Giampapa}, \& {Schmitt}}{{Fleming}
  et~al.}{2000}]{fgs00}
{Fleming}, T.~A., {Giampapa}, M.~S.,  \& {Schmitt}, J.~H.~M.~M. 2000, \apj,
  533, 372

\bibitem[\protect\citeauthoryear{{Gizis} et~al.}{{Gizis} et~al.}{2000}]{gmr+00}
{Gizis}, J.~E., {Monet}, D.~G., {Reid}, I.~N., {Kirkpatrick}, J.~D., {Liebert},
  J.,  \& {Williams}, R.~J. 2000, \aj, 120, 1085

\bibitem[\protect\citeauthoryear{{G{\"u}del}}{{G{\"u}del}}{2002}]{gud02}
{G{\"u}del}, M. 2002, \araa, 40, 217

\bibitem[\protect\citeauthoryear{{G{\"u}del}}{{G{\"u}del}}{2004}]{gud04}
{G{\"u}del}, M. 2004, \aapr, 12, 71

\bibitem[\protect\citeauthoryear{{Guedel} \& {Benz}}{{Guedel} \&
  {Benz}}{1993}]{gb93}
{Guedel}, M.,  \& {Benz}, A.~O. 1993, ApJ, 405, L63

\bibitem[\protect\citeauthoryear{{Hallinan} et~al.}{{Hallinan}
  et~al.}{2007}]{hbl+07}
{Hallinan}, G., et~al. 2007, \apjl, 663, L25

\bibitem[\protect\citeauthoryear{{Hawley}, {Gizis}, \& {Reid}}{{Hawley}
  et~al.}{1996}]{hgr96}
{Hawley}, S.~L., {Gizis}, J.~E.,  \& {Reid}, I.~N. 1996, \aj, 112, 2799

\bibitem[\protect\citeauthoryear{{Hawley} \& {Johns-Krull}}{{Hawley} \&
  {Johns-Krull}}{2003}]{hj03}
{Hawley}, S.~L.,  \& {Johns-Krull}, C.~M. 2003, \apjl, 588, L109

\bibitem[\protect\citeauthoryear{{Hook} et~al.}{{Hook} et~al.}{2004}]{hja+04}
{Hook}, I.~M., {J{\o}rgensen}, I., {Allington-Smith}, J.~R., {Davies}, R.~L.,
  {Metcalfe}, N., {Murowinski}, R.~G.,  \& {Crampton}, D. 2004, \pasp, 116, 425

\bibitem[\protect\citeauthoryear{{James} et~al.}{{James} et~al.}{2000}]{jjj+00}
{James}, D.~J., {Jardine}, M.~M., {Jeffries}, R.~D., {Randich}, S., {Collier
  Cameron}, A.,  \& {Ferreira}, M. 2000, \mnras, 318, 1217

\bibitem[\protect\citeauthoryear{{Jardine} \& {Unruh}}{{Jardine} \&
  {Unruh}}{1999}]{ju99}
{Jardine}, M.,  \& {Unruh}, Y.~C. 1999, \aap, 346, 883

\bibitem[\protect\citeauthoryear{{Kraft}}{{Kraft}}{1967}]{kra67}
{Kraft}, R.~P. 1967, \apj, 150, 551

\bibitem[\protect\citeauthoryear{{Krishnamurthi}, {Leto}, \&
  {Linsky}}{{Krishnamurthi} et~al.}{1999}]{kll99}
{Krishnamurthi}, A., {Leto}, G.,  \& {Linsky}, J.~L. 1999, \aj, 118, 1369

\bibitem[\protect\citeauthoryear{{Leto} et~al.}{{Leto} et~al.}{2000}]{lpl+00}
{Leto}, G., {Pagano}, I., {Linsky}, J.~L., {Rodon{\`o}}, M.,  \& {Umana}, G.
  2000, \aap, 359, 1035

\bibitem[\protect\citeauthoryear{{Liebert} et~al.}{{Liebert}
  et~al.}{2003}]{lkc+03}
{Liebert}, J., {Kirkpatrick}, J.~D., {Cruz}, K.~L., {Reid}, I.~N., {Burgasser},
  A., {Tinney}, C.~G.,  \& {Gizis}, J.~E. 2003, \aj, 125, 343

\bibitem[\protect\citeauthoryear{{Linsky} et~al.}{{Linsky}
  et~al.}{1995}]{lwb+95}
{Linsky}, J.~L., {Wood}, B.~E., {Brown}, A., {Giampapa}, M.~S.,  \&
  {Ambruster}, C. 1995, \apj, 455, 670

\bibitem[\protect\citeauthoryear{{Mart{\'{\i}}n} et~al.}{{Mart{\'{\i}}n}
  et~al.}{1999}]{mdb+99}
{Mart{\'{\i}}n}, E.~L., {Delfosse}, X., {Basri}, G., {Goldman}, B.,
  {Forveille}, T.,  \& {Zapatero Osorio}, M.~R. 1999, \aj, 118, 2466

\bibitem[\protect\citeauthoryear{{Mohanty} \& {Basri}}{{Mohanty} \&
  {Basri}}{2003}]{mb03}
{Mohanty}, S.,  \& {Basri}, G. 2003, \apj, 583, 451

\bibitem[\protect\citeauthoryear{{Narain} \& {Ulmschneider}}{{Narain} 
\& {Ulmschneider}}{1996}]{nu96}
{Narain}, U.,  \& {Ulmschneider}, P. 1996, Space Sci.~Rev., 75, 453

\bibitem[\protect\citeauthoryear{{Neupert}}{{Neupert}}{1968}]{neu68}
{Neupert}, W.~M. 1968, \apjl, 153, L59

\bibitem[\protect\citeauthoryear{{Osten} et~al.}{{Osten}
  et~al.}{2006a}]{oha+06}
{Osten}, R.~A., {Hawley}, S.~L., {Allred}, J., {Johns-Krull}, C.~M., {Brown},
  A.,  \& {Harper}, G.~M. 2006a, \apj, 647, 1349

\bibitem[\protect\citeauthoryear{{Osten} et~al.}{{Osten}
  et~al.}{2006b}]{ohb+06}
{Osten}, R.~A., {Hawley}, S.~L., {Bastian}, T.~S.,  \& {Reid}, I.~N. 2006b,
  \apj, 637, 518

\bibitem[\protect\citeauthoryear{{Pallavicini} et~al.}{{Pallavicini}
  et~al.}{1981}]{pgr+81}
{Pallavicini}, R., {Golub}, L., {Rosner}, R., {Vaiana}, G.~S., {Ayres}, T.,  \&
  {Linsky}, J.~L. 1981, \apj, 248, 279

\bibitem[\protect\citeauthoryear{{Parker}}{{Parker}}{1955}]{par55}
{Parker}, E.~N. 1955, \apj, 122, 293

\bibitem[\protect\citeauthoryear{{Patten} et~al.}{{Patten}
  et~al.}{2006}]{psb+06}
{Patten}, B.~M., et~al. 2006, \apj, 651, 502

\bibitem[\protect\citeauthoryear{{Phan-Bao} et~al.}{{Phan-Bao}
  et~al.}{2007}]{pol+07}
{Phan-Bao}, N., {Osten}, R.~A., {Lim}, J., {Mart{\'{\i}}n}, E.~L.,  \& {Ho},
  P.~T.~P. 2007, \apj, 658, 553

\bibitem[\protect\citeauthoryear{{Pizzolato} et~al.}{{Pizzolato}
  et~al.}{2003}]{pmm+03}
{Pizzolato}, N., {Maggio}, A., {Micela}, G., {Sciortino}, S.,  \& {Ventura}, P.
  2003, \aap, 397, 147

\bibitem[\protect\citeauthoryear{{Reid} et~al.}{{Reid} et~al.}{2003}]{rcl+03}
{Reid}, I.~N., et~al. 2003, \aj, 125, 354

\bibitem[\protect\citeauthoryear{{Reid} et~al.}{{Reid} et~al.}{1999}]{rkg+99}
{Reid}, I.~N., {Kirkpatrick}, J.~D., {Gizis}, J.~E.,  \& {Liebert}, J. 1999,
  \apjl, 527, L105

\bibitem[\protect\citeauthoryear{{Reiners} \& {Basri}}{{Reiners} \&
  {Basri}}{2007}]{rb07}
{Reiners}, A.,  \& {Basri}, G. 2007, \apj, 656, 1121

\bibitem[\protect\citeauthoryear{{Rutledge} et~al.}{{Rutledge}
  et~al.}{2000}]{rbm+00}
{Rutledge}, R.~E., {Basri}, G., {Mart{\'i}n}, E.~L.,  \& {Bildsten}, L. 2000,
  ApJ, 538, L141

\bibitem[\protect\citeauthoryear{{St{\c e}pie{\'n}}, {Schmitt}, \&
  {Voges}}{{St{\c e}pie{\'n}} et~al.}{2001}]{ssv01}
{St{\c e}pie{\'n}}, K., {Schmitt}, J.~H.~M.~M.,  \& {Voges}, W. 2001, \aap,
  370, 157

\bibitem[\protect\citeauthoryear{{Stelzer} et~al.}{{Stelzer}
  et~al.}{2006}]{ssm+06}
{Stelzer}, B., {Schmitt}, J.~H.~M.~M., {Micela}, G.,  \& {Liefke}, C. 2006,
  \aap, 460, L35

\bibitem[\protect\citeauthoryear{{Stewart} et~al.}{{Stewart}
  et~al.}{1988}]{sis+88}
{Stewart}, R.~T., {Innis}, J.~L., {Slee}, O.~B., {Nelson}, G.~J.,  \& {Wright},
  A.~E. 1988, \aj, 96, 371

\bibitem[\protect\citeauthoryear{{Tinney}}{{Tinney}}{1996}]{tin96}
{Tinney}, C.~G. 1996, \mnras, 281, 644

\bibitem[\protect\citeauthoryear{{van den Besselaar} et~al.}{{van den
  Besselaar} et~al.}{2003}]{vrm+03}
{van den Besselaar}, E.~J.~M., {Raassen}, A.~J.~J., {Mewe}, R., {van der Meer},
  R.~L.~J., {G{\"u}del}, M.,  \& {Audard}, M. 2003, \aap, 411, 587

\bibitem[\protect\citeauthoryear{{Vilhu}}{{Vilhu}}{1984}]{vil84}
{Vilhu}, O. 1984, \aap, 133, 117

\bibitem[\protect\citeauthoryear{{Vilhu} \& {Walter}}{{Vilhu} \&
  {Walter}}{1987}]{vw87}
{Vilhu}, O.,  \& {Walter}, F.~M. 1987, \apj, 321, 958

\bibitem[\protect\citeauthoryear{{Walkowicz}, {Hawley}, \& {West}}{{Walkowicz}
  et~al.}{2004}]{whw04}
{Walkowicz}, L.~M., {Hawley}, S.~L.,  \& {West}, A.~A. 2004, \pasp, 116, 1105

\bibitem[\protect\citeauthoryear{{West} et~al.}{{West} et~al.}{2004}]{whw+04}
{West}, A.~A., et~al. 2004, \aj, 128, 426

\bibitem[\protect\citeauthoryear{{White}, {Jackson}, \& {Kundu}}{{White}
  et~al.}{1989}]{wjk89}
{White}, S.~M., {Jackson}, P.~D.,  \& {Kundu}, M.~R. 1989, \apjs, 71, 895


\end{thebibliography}
\end{document}